\documentclass[prx,reprint,notitlepage,amsmath,amssymb,amsfonts,longbibliography]{revtex4-1}
\usepackage{graphicx}% Include figure files
\usepackage{dcolumn}% Align table columns on decimal point
\usepackage{bm}% bold math
\usepackage{float}
\usepackage{color}
\usepackage{refstyle}
\usepackage{mathrsfs}
\usepackage{esint}
\usepackage{enumitem}
\usepackage{multirow}
\usepackage[unicode=true,pdfusetitle,bookmarks=true,
bookmarksnumbered=false,bookmarksopen=false,breaklinks=false,
pdfborder={0 0 0},backref=false,colorlinks=true,citecolor=red]
{hyperref}
\usepackage{cleveref}
\providecommand\eqref[1]{\ref{eq:#1}}
\renewcommand\b[1]{{\bf  #1}}
\renewcommand\vec[1]{\boldsymbol{#1}}
\newcommand\vphi{\varphi}

\newcommand\del{\nabla}
\newcommand\e{\epsilon}
\newcommand\vare{\varepsilon}
\newcommand\dd{\mathrm{d}}
\allowdisplaybreaks[3]
 
\numberwithin{equation}{section}

\begin{document}

\title{Thermalized buckling of isotropically compressed thin sheets}
\author{Suraj Shankar$^1$ and David R.~Nelson$^{1,2}$}
\affiliation{
$^1$Department of Physics, Harvard University, Cambridge, MA 02138, USA\\
$^2$Department of Molecular and Cellular Biology and School of Engineering and Applied Sciences, Harvard University, Cambridge, Massachusetts 02138, USA}
%\email[]{suraj\_shankar@fas.harvard.edu}

\date{\today}
\begin{abstract}
	The buckling of thin elastic sheets is a classic mechanical instability that occurs over a wide range of scales. In the extreme limit of atomically thin membranes like graphene, thermal fluctuations can dramatically modify such mechanical instabilities. We investigate here the delicate interplay of boundary conditions, nonlinear mechanics and thermal fluctuations in controlling buckling of confined thin sheets under isotropic compression. We identify two inequivalent mechanical ensembles based on the boundaries at constant strain (isometric) or at constant stress (isotensional) conditions. Remarkably, in the isometric ensemble, boundary conditions induce a novel long-ranged nonlinear interaction between the local tilt of the surface at distant points. This interaction combined with a spontaneously generated thermal tension leads to a renormalization group description of two distinct universality classes for thermalized buckling, realizing a mechanical variant of Fisher-renormalized critical exponents. We formulate a complete scaling theory of buckling as an unusual phase transition with a size dependent critical point, and discuss experimental ramifications for the mechanical manipulation of ultrathin nanomaterials.
\end{abstract}
\maketitle
\section{Introduction}
Thin sheets with a resistance to shear can accommodate compressive stresses through an array of mechanical instabilities, including buckling \cite{LandauElasticityBook,koiter1967stability}, wrinkling \cite{Cerda2002,*Cerda2003,Davidovitch11,Vandeparre11}, folding \cite{pocivavsek2008stress,diamant2011compression} and crumpling \cite{benamar97,*Cerda1998,*Cerda1999,witten2007stress}, all controlled essentially by geometry. Although once disregarded as undesirable modes of failure, instabilities now play a central role in the design of mechanical metamaterials \cite{krieger2012buckling,bertoldi2017flexible} as they combine complex morphologies with mechanical functionality. In recent years, rapid miniaturization has driven intense research efforts in developing similar metamaterials on a much smaller scale \cite{Blees2015,rogers2016origami,*xu2017ultrathin,*miskin2018graphene,*reynolds2019capillary,*grosso2020graphene}. In this regard, atomically thin two dimensional (2D) materials such as graphene, MoS$_2$ or BN \cite{novoselov2005two,katsnelson2012graphene} are particularly promising and offer unprecedented opportunities to study classical elasticity and mechanics in the ultimate limit in thin sheets, where thermal fluctuations can play a dominant role \cite{nelson2004statistical,katsnelson2012graphene}.

In such ultrathin flexible materials thermal fluctuations can dramatically renormalize the mechanical properties in a scale-dependent fashion \cite{nelson1987fluctuations}. Out of plane (flexural) deformations allow tensionless solid membranes to exhibit a remarkable thermally wrinkled, yet flat phase with a scale-dependent bending rigidity and strongly softened elastic moduli \cite{bowick2001statistical,nelson2004statistical}. While thin sheets favour bending over energetically expensive stretching, geometry links the two as any bending-induced Gaussian curvature inevitably causes stretching as well. This basic feature underlies many of the impressive finite temperature properties. Nanoindentation measurements in graphene \cite{lee2008measurement} and MoS$_2$ \cite{bertolazzi2011stretching} monolayers yield exceptionally high Young's moduli on the nanoscale as expected from strong covalent bonding. Yet on larger scales $\sim10~\mu$m, recent experiments with freely suspended graphene have demonstrated a $\sim4000$ fold enhancement of the bending rigidity \cite{Blees2015} and a factor $\sim20$ reduction in the in-plane stiffness \cite{nicholl2015effect,*nicholl2017hidden}, due to a combination of thermally generated and static ripples \cite{meyer2007structure,kovsmrlj2013mechanical,xu2014unusual}, highlighting the importance of flexural fluctuations.

While the anomalous mechanics of thermalized membranes has been extensively explored, the role of confinement and boundaries is much less appreciated. Supported or clamped edges are one of the most commonly encountered boundary conditions, in electromechanical resonators \cite{bunch2007electromechanical,chen2009performance}, multistable switches \cite{loh2012nanoelectromechanical} and in nanomechanical devices \cite{Blees2015,nicholl2015effect,nicholl2017hidden}. Geometric confinement at the boundary can induce prestrains in the sample that can cause large scale instabilities such as wrinkling \cite{bao2009controlled}. As a result, in recent years, exploring the influence of external stresses on the mechanics of fluctuating membranes has been a topic of prime interest \cite{amorim2016novel}. While there has been some theoretical work, both old \cite{guitter1988crumpling,*guitter1989thermodynamical} and new \cite{roldan2011suppression,kovsmrlj2016response,bonilla2016critical,burmistrov2018stress,*burmistrov2018differential,*saykin2020absolute}, complemented by more recent large scale numerical simulations \cite{jiang2014buckling,bowick2017non,wan2017thermal,sgouros2018compressive,morshedifard2020buckling,hanakata2020thermal}, elucidating the role of boundaries in controlling the nonlinear mechanics and buckling of thermalized sheets, particularly for compressions which attempt to impose a nonzero Gaussian curvature, remains a challenging problem.

Motivated by the above, in this paper, we pose and answer the following question - what is the finite temperature version of the buckling transition in an isotropically compressed thin sheet? Euler buckling represents the simplest mechanical instability a thin elastic body can undergo and it provides an attractive setting to investigate the interplay of thermal fluctuations and boundary conditions along with the geometric nonlinearities inherent to thin plate mechanics. In particular, we focus on the universal aspects of the transition such as critical scaling exponents that are independent of microscopic details. By combining a full renormalization group analysis along with a general scaling theory, we provide a complete description of thermalized buckling as a genuine phase transition that exhibits critical scaling along with more unusual features such as a sensitive dependence on system size and the choice of boundary conditions. Apart from new statistical mechanical results, the scaling framework we propose also yields key predictions that have important ramifications for experiments that we highlight below.

In the rest of the introduction, we summarize our main results and outline the structure of the paper.

\subsection{Results and outline}
A key outcome of our work is a renormalization group analysis, augmented by a scaling description of buckling in isotropically compressed thermalized thin sheets. There are two main reasons why finite temperature buckling, when viewed as a phase transition, is distinct from conventional critical phenomena. The first is that buckling is strongly size dependent, even at zero temperature by virtue of it being a long-wavelength instability \cite{LandauElasticityBook}. The second is the remarkable fact that freely fluctuating elastic sheets exhibit a flat phase \cite{nelson1987fluctuations} with critical fluctuations over an extended range of temperatures. Both these features are characteristic of thin sheets, arising from an interplay of geometry and mechanics, and form the basis of our results below.

{
\renewcommand{\thesubsection}{\arabic{subsection}}
\setcounter{subsection}{0}
\subsection{\emph{Ensemble inequivalence \& Fisher renormalization}}
\begin{figure}[]
	\centering{\includegraphics[width=0.45\textwidth]{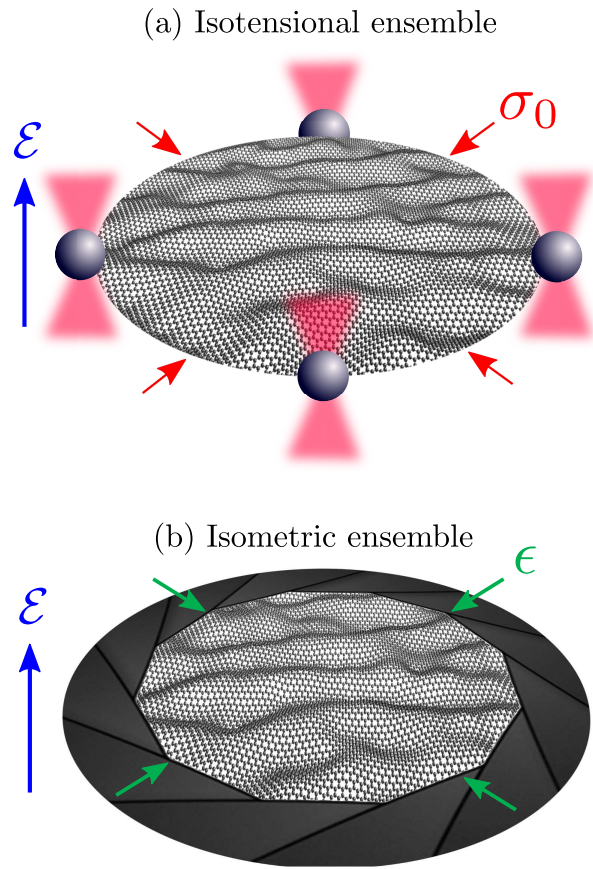}}
	\caption{A sketch showing a possible realization of the two mechanical ensembles, for example, in an atomically thin sheet of graphene. (a) In the isotensional ensemble, a constant inward external stress $\sigma_0$ is applied to the membrane, while the boundary displacement fluctuates. This set-up might be realized in the same fashion as in single molecule experiments, by using feedback-controlled multiplexed optical tweezers to actuate under constant force conditions, similar to experiments recently used to probe the mechanical response of red blood cells \cite{turlier2016equilibrium}. (b) The isometric ensemble instead corresponds to a clamped boundary with the external load imposed via a global strain $\e$. While, current experiments with graphene typically suspend monolayers across fixed size holes \cite{nicholl2015effect}, a variable aperture size tuned by a camera shutter mechanism could be used to tune the strain isotropically. An external symmetry breaking field $\mathcal{E}$ perpendicular to average plane of the sheet can also be applied in either ensemble to bias the direction of buckling.}
	\label{fig:ensemble}
\end{figure}

Thin sheets can be loaded in-plane either by prescribing the external strain (isometric) or the external stress (isotensional) at the boundary (see Fig.~\ref{fig:ensemble} for an illustration). These constitute dual mechanical ensembles, in analogy with thermodynamic ensembles \cite{gibbs1902elementary}. While it is well known that boundary conditions can modify nonuniversal quantities such as the buckling threshold \cite{LandauElasticityBook}, we discover that universal scaling exponents (defined in Sec.~\ref{sec:exp}) can also exhibit a remarkable sensitivity to boundary conditions! We demonstrate this fact explicitly within a systematic $\vare=4-D$ expansion for a general $D$-dimensional solid embedded in $d>D$ dimensional space (discussed in Secs.~\ref{sec:rg} and~\ref{sec:buckling} with details relegated to Appendix~\ref{app:rgd}) along with a simpler, but uncontrolled, one-loop calculation performed directly in the physical dimensions of $D=2$ and $d=3$ in Appendix~\ref{app:rg}. Our calculations show that buckling in the two mechanical ensembles is in fact controlled by two distinct fixed points, with different scaling exponents that are summarized in Table~\ref{tab:exponents}.
This remarkable departure from conventional wisdom demonstrates the nonequivalence of mechanical ensembles in thermally fluctuating thin sheets and highlights the subtle nature of membrane statistical mechanics. Thus, in the simplest setting of isotropic compression, we find thermalized plate buckling is described by two distinct critical points characterizing the isometric and isotensional universality classes that are distinguished simply by the boundary conditions imposed.

Although surprising, the inequivalence of ensembles has precedence in critical phenomena. A critical point engenders fluctuations on all scales which under appropriate conditions can result in scaling exponents that change upon switching to a dual or constrained ensemble. This is known as Fisher renormalization \cite{fisher1968renormalization,*lipa1968critical,*fisher1970visibility}. In our case, however, by tuning to the buckling threshold, we approach the flat phase of a free-standing membrane which characterizes an entire low temperature \emph{critical phase} with scale invariant fluctuations! Furthermore, we find that the imposition of a fixed strain (in contrast to a fixed stress) boundary condition induces a novel long-ranged nonlinearity that couples the local \emph{tilt} of the surface at far away points, which we derive in Sec.~\ref{sec:ensemble} and Appendix~\ref{app:iso} through a careful consideration of zero modes and boundary conditions. This nonlocal term, which can also be important far from the buckling transition, softly enforces the geometric constraint of global inextensibilitiy, which simultaneously shifts the buckling threshold by a spontaneously generated thermal tension and modifies the critical exponents via a mechanical variant of Fisher renormalization.

\subsection{\emph{Size-dependent scaling theory}}
The long-wavelength nature of the buckling instability endows it with both a system size dependent threshold and a macroscopic mechanical response \cite{LandauElasticityBook}, features that are retained even at finite temperature. This size dependence is unusual though, from the point of view of critical phenomena, and behaves as a \emph{dangerously irrelevant} variable \cite{amit1982dangerous,*Gunton1973RenormalizationGI,*nelson1976coexistence} that modifies scaling exponents in nontrivial ways. As a result, we derive new exponent identites in Sec.~\ref{sec:scaling} that relate different scaling exponents in both ensembles, mirroring classic results from conventional critical phenomena \cite{goldenfeld2018lectures}. Many of these relations are also summarized in Table~\ref{tab:exponents}. By combining scaling with general thermodynamic arguments, we also explicitly demonstrate how buckling physics in both ensembles is a mechanical variant of Fisher renormalization. Note that, the construction of a consistent scaling theory for thermalized buckling is a significant achievement as it not only clarifies previous confounding results \cite{guitter1988crumpling,guitter1989thermodynamical,kovsmrlj2016response} by correctly accounting for nontrivial system size dependence and ensemble inequivalence, but it also yields a unified framework that incorporates experimentally relevant boundary conditions and symmetry breaking fields.

\subsection{\emph{Experimental consequences}}
Our work illustrates the spectacular ways in which geometry, boundary effects and thermal fluctuations can conspire to produce unexpected phenomena, and suggests that extending thin body mechanics to finite temperature is a rich and challenging enterprise, requiring great care. Given the ubiquity and ease of manipulating strain rather than stress in experiments, our results have important implications for the rational design of strain engineered nanodevices and interpretation of mechanical measurements in ultrathin materials in the presence of thermal fluctuations.

\begin{figure}[]
	\centering{\includegraphics[width=0.5\textwidth]{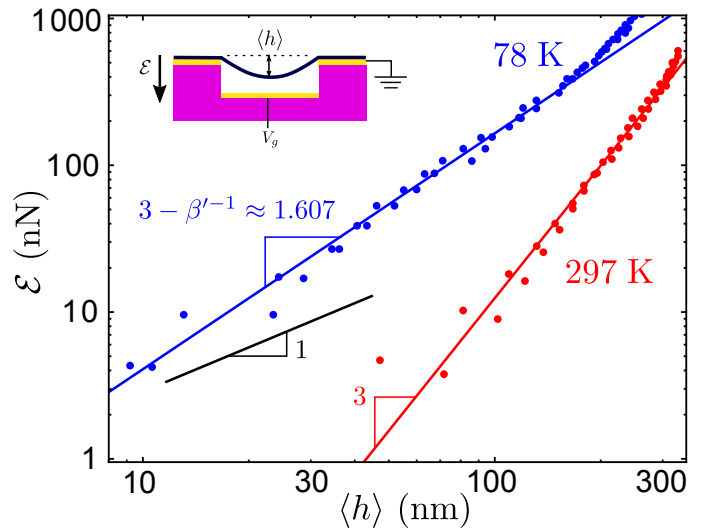}}
	\caption{Experimentally measured height response curve of a clamped graphene sheet suspended over a circular hole of radius $6.2~\mu$m in a cryostat at two different temperatures, $T=78~$K (blue dots) and $T=297~$K (red dots). The data is reproduced from Ref.~\cite{storch2018young}. The electrically integrated graphene devices are capacitively actuated out of the plane using an electrostatic force $\mathcal{E}\propto V_g^2$ ($V_g$ is the gate voltage) and the average deflection $\langle h\rangle$ of the sheet is measured using laser interferometry, see inset for a sketch of the setup and Ref.~\cite{storch2018young} for further details. The red and blue lines are guides to the eye showing the exponent of the nonlinear force response. While the $T=297~$K data (red dots) shows a strong cubic dependence on the height, the $T=78~$K data (blue dots) shows a smaller slope that matches well with our theoretical prediction using the \emph{isometric} ensemble exponent $3-1/\beta'\approx1.607$. Note that, this slope is significantly different from a slope of unity (black line), as would be predicted in both the \emph{isotensional} ensemble and within mean field theory.}
	\label{fig:data}
\end{figure}

A common setup to probe the mechanical properties of graphene involves measuring the force-displacement curve for a clamped monolayers that is deformed by the application of stresses or external fields. As a simple example, we shall focus here on a simple circular geometry with sheets draped across holes and deflected by a normal electric field ($\mathcal{E}$) as employed in recent experiments \cite{nicholl2015effect,storch2018young} (see inset in Fig.~\ref{fig:data} for a sketch). Other geometries including ribbons cantilevered at an edge \cite{Blees2015} or suspended across trenches \cite{bunch2007electromechanical,chen2009performance,bao2009controlled} are also possible, but we don't address them here. A crucial ingredient in the interpretation of force response measurements in these devices is a mechanical ``equation of state'' that relates the externally applied field $\mathcal{E}$ and the in-plane tension $\Delta\sigma$ to the magnitude of the average deflection of the sheet $\langle h\rangle$. A zero-temperature mean-field description following classical elasticity (detailed in Sec.~\ref{sec:mft}) gives
\begin{equation}
	\mathcal{E}=c_1\dfrac{\Delta\sigma}{R^2}\langle h\rangle+c_2\dfrac{Y}{R^4}\langle h\rangle^3\;,\label{eq:mfteos}
\end{equation}
where $c_{1,2}$ are calculable numerical constants, $Y$ is the sought after 2D Young's modulus and $R$ is the size of the sheet. Importantly, when the sheet is constrained at fixed stress $\Delta\sigma$, i.e., the isotensional ensemble, Eq.~\ref{eq:mfteos} applies exactly even at finite temperature upon simply replacing $Y$ by its scale dependent renormalized value (see Sec.~\ref{sec:scaling} for details).

However, nearly all the force measurement experiments are instead conducted with fixed strain and clamped boundary conditions, i.e., in the \emph{isometric} ensemble. Our general scaling theory and renormalization group analysis provides a different equation
\begin{equation}
	\mathcal{E}=c_1\dfrac{\Delta\sigma}{R^2}\left(\dfrac{Y}{R^2}\right)^{1-1/2\beta'}\langle h\rangle^{3-1/\beta'}+c_2\dfrac{Y}{R^4}\langle h\rangle^3\;,\label{eq:isoeos}
\end{equation}
where the tension $\Delta\sigma=B\Delta\e$ is now given by the 2D bulk modulus $B$ and the imposed in-plane strain $\Delta\e$. Strikingly, Eq.~\ref{eq:isoeos} involves a new order parameter exponent $\beta'\approx0.718$ that controls the asymptotic nonlinear force response. The difference between Eq.~\ref{eq:mfteos} and Eq.~\ref{eq:isoeos} makes it clear that using Eq.~\ref{eq:mfteos} for clamped sheets, as is conventionally done, can lead to significantly wrong results. In fact, recent experimental measurements \cite{storch2018young} on graphene drumheads match well with our theoretical predictions, see Fig.~\ref{fig:data}. The strain in the sample increases upon lowering temperature \cite{storch2018young,chen2009performance}, notwithstanding the theoretically expected negative thermal expansion coefficient of graphene \cite{kovsmrlj2016response,bao2009controlled}, presumably due to surface contaminants. As a result, while the classical cubic response dominates at higher temperature with weak tension (red dots in Fig.~\ref{fig:data}), the anomalous nonlinear response with $\mathcal{E}\propto\langle h\rangle^{1.607}$ (blue dots in Fig.~\ref{fig:data}) emerges for higher strains at lower temperature. A systematic analysis exploring how this nonlinear response affects the extraction of the Young's modulus is left for future work.

This result highlights the direct relevance of our work to the correct interpretation of mechanical measurements in graphene devices, not only for the circular geometries studied here, but also cantilevers and doubly clamped ribbons. We believe that recognizing the fundamental distinction between the isotensional and the isometric ensembles is essential to such endeavours. While Fig.~\ref{fig:data} depicts a static example, a dynamical extension of Eq.~\ref{eq:isoeos} including dissipation and inertia along with a time-varying field $\mathcal{E}(t)$ provides a simple description of periodicaly driven electromechanical oscillators \cite{lifshitz2008nonlinear}. Although a full dynamical analysis is beyond the scope of this current work, we can already appreciate the presence of a strong nonlinear response (Eq.~\ref{eq:isoeos}) in the small deflection limit, which allows for higher quality factors and bistability \cite{lifshitz2008nonlinear} even for a weak drive. Note that, such an anomalous response is only elicited in the isometric ensemble, emphasizing once again the importance of boundary conditions.

Although we focus on isotropic compression, we expect suitable extensions of this framework to be applicable to recent numerical simulations \cite{morshedifard2020buckling,hanakata2020thermal} and experiments \cite{Blees2015,bao2009controlled,xu2014unusual} on compressed ribbons that have begun addressing the anomalous mechanics of \emph{anisotropic} buckling in constrained sheets. While much of our discussion has been based on atomic crystals, we emphasize our results are generic and also apply to organic 2D polymers, such as naturally occuring DNA kinetoplasts \cite{klotz2020equilibrium}, the spectrin cytoskeleton \cite{schmidt1993existence} and reconstituted spider silk \cite{hermanson2007engineered}, or synthetic molecular crystals \cite{servalli2018synthesizing,*payamyar2016two} and possibly polymerosomes \cite{shum2008microfluidic} with very large radii. In Sec.~\ref{sec:discussion}, we conclude with a brief discussion of the broader significance of our results to experiments on atomically thin crystalline membranes and possible future directions.

%The paper is structured as follows. We first introduce and briefly review the elastic theory of thin plates in Sec.~\ref{sec:plate}, following which we formulate its statistical mechanics as a continuum field theory in Sec.~\ref{sec:ensemble}. Here we eliminate the in-plane strains in favour of out-of-plane deformations to obtain a reduced free energy for a fluctuating sheet. In doing so, we pay special attention to the relevant boundary conditions, which allows us to identify the two mechanical ensembles (isometric and isotensional) that characterize the mode of external loading on the sheet.
%Given our focus on buckling, we next define the various scaling exponents of interest in Sec.~\ref{sec:exp}, just as in conventional critical phenomena \cite{stanley1987introduction,goldenfeld2018lectures}.
%With the notation in place, our first step of analysis is a simple mean-field theory in Sec.~\ref{sec:mft}, where we recapitulate classic results of plate buckling in the language of phase transitions (for a recapitualtion of buckling transitions for ribbons subject to uniaxial compression, see Ref.~\cite{hanakata2020thermal}). Thermal fluctuations are included within a simple Gaussian framework in Sec.~\ref{sec:gaussian} which allows us to assess when geometric nonlinearities are important.
}

\section{Thin plate elasticity}
\label{sec:plate}
Here, we focus on the physically relevant continuum elastic description of a thin 2D sheet fluctuating in 3D space (a generalization for a $D$-dimensional solid embedded in $d$-dimensional ($d>D$) Euclidean space, useful for certain calculations, is provided in Appendix~\ref{app:rgd}).
The deformation of a thin flat sheet is parametrized in the Monge gauge using a 2D in-plane displacement vector $\b{u}$ and a height field $h$. The total elastic energy of the sheet involves both stretching and bending contributions and is given by \cite{LandauElasticityBook}
\begin{equation}
	\mathcal{H}=\int\dd^2r\left[\dfrac{\kappa}{2}(\del^2h)^2+\mu u_{ij}^2+\dfrac{\lambda}{2}u_{kk}^2-\mathcal{E}h\right]-\oint_{\mathcal{C}}\dd\ell\;\hat{\nu}_i\sigma^{\rm ext}_{ij}u_j\;.\label{eq:H}
\end{equation}
The Lam{\'e} parameters are $\mu$ and $\lambda$, and $\kappa$ is the bending rigidity. The final boundary integral is the work done by an external stress $\vec{\sigma}^{\rm{ext}}$ with $\hat{\vec{\nu}}$ being the outward normal (within the plane) to the boundary curve $\mathcal{C}$. The penultimate term corresponds to the potential energy due to an external out of plane field $\mathcal{E}$ which couples directly to the height of the membrane. Such a perturbation can be realized by an electric field $E$, with $\mathcal{E}=\rho_{q}E$, where $\rho_q$ is the electric charge density on the surface, while in the presence of gravity, we have $\mathcal{E}=\rho_mg$, where $\rho_m$ is the mass density and $g$ is the gravitational acceleration. The strain tensor
\begin{equation}
	u_{ij}=\dfrac{1}{2}\left(\partial_iu_j+\partial_ju_i+\partial_ih\partial_jh\right)\;,
\end{equation}
encodes the geometric nonlinearity inherent to thin sheets.
We neglect higher order terms in the in-plane displacements which are small and irrelevant on large scales for a thin sheet \cite{LandauElasticityBook}. As an aside, note that, for a $D$-dimensional manifold embedded in $d$-dimensional space, the displacement field $\b{u}$ has $D$ components and the height field is no longer a scalar, but instead a vector $\b{h}$ with codimension $d_c=d-D>0$ components. The nonlinear strain tensor in this case is then $u_{ij}=(\partial_iu_j+\partial_ju_i+\partial_i\b{h}\cdot\partial_j\b{h})/2$. The relative importance of stretching versus bending energies is captured by a dimensionless F{\"o}ppl-von K{\'a}rm{\'a}n number, which in 2D is given by
\begin{equation}
	{\rm vK}=\dfrac{YR^2}{\kappa}\;,
\end{equation}
where $Y=4\mu(\mu+\lambda)/(2\mu+\lambda)$ is the 2D Young's modulus and $R$ is a characteristic linear dimension of the sheet. When viewing the sheet as thin slice of a bulk elastic material, its bending modulus and stiffness are related as $\kappa=Y_{\rm 3D}t^3/[12(1-\nu_{\rm 3D}^2)]$ and $Y=Y_{\rm 3D}t$, where $Y_{\rm 3D}$ is the 3D Young's modulus, $\nu_{\rm 3D}$ the 3D Poisson's ratio and $t$ is the thickness of the sheet \cite{audoly2010elasticity}. As a result, ${\rm vK}\sim(R/t)^2$ is essentially controlled by geometry with ${\rm vK}\gg 1$ for a thin sheet ($t/R\ll 1$), reflecting the dominance of geometrically nonlinear isometric deformations, i.e., bending without stretching. An ordinary sheet of paper has a ${\rm vK}\approx 10^6$ while a $1~\mu$m size graphene monolayer has a microscopic ${\rm vK}\approx 10^9$ (using the atomic scale values for $\kappa\approx1.1~$eV \cite{katsnelson2012graphene} and $Y\approx 340~$N/m \cite{lee2008measurement}).
\section{Mechanical ensembles}
\label{sec:ensemble}
The properties of a thermalized elatic membrane at temperature $T$ are computed through the equilibrium partition function,
\begin{equation}
	\mathcal{Z}=\int\mathcal{D}h\mathcal{D}\b{u}\;e^{-\mathcal{H}/k_BT}\;,
\end{equation}
where $k_B$ is the Boltzmann constant. As the in-plane phonons ($\b{u}$) only appear quadratically in $\mathcal{H}$ they can be integrated out exactly to give an effective free energy $\mathcal{F}=-k_BT\ln\int\mathcal{D}\b{u}\;e^{-\mathcal{H}/k_BT}$. To do this, we separate out the average strain and Fourier transform the nonzero wavelength deformations. While the calculation for the wavevector $\b{q}\neq\b{0}$ modes is standard \cite{nelson2004statistical}, the homogeneous $\b{q}=\b{0}$ strain mode needs to be handled with care in the presence of various boundary conditions. We shall focus on isotropic loading at the boundary and neglect external shear or compressional loading of ribbons for simplicity. This leaves us with two possibilities, which are the
\begin{enumerate}[label=(\roman*)]
	\item Fixed stress or \emph{isotensional} ensemble, and the
	\item Fixed strain or \emph{isometric} ensemble.
\end{enumerate}
Note the latter could be realized with clamped circular boundary conditions.
Stress and strain (equivalently, force and displacement) are thermodynamically conjugate variables and the elliptic nature of elasticity prohibits specifying both at a boundary simultaneously. Hence, we have two natural mechanical ensembles akin to the isobaric $(N,P,T)$ and isochoric $(N,V,T)$ ensembles of statistical mechanics respectively, that are dual to each other. In Fig.~\ref{fig:ensemble}, we sketch a possible realization of the two mechanical ensembles in an atomically or molecularly thin suspended sheet.

In the isotensional ensemble, the sheet is driven by an external isotropic stress
\begin{equation}
	\sigma^{\rm ext}_{ij}=\sigma_0\delta_{ij}\;,
\end{equation}
with no further constraints on the zero modes of the displacement or strain. As a result, the boundary can freely displace in the plane under the action of $\sigma_0\neq 0$. Note that $\sigma_0>0$ corresponds to a tensile stress while $\sigma_0<0$ is compressive stress, a situation studied in Ref.~\cite{kovsmrlj2016response}. In the isometric ensemble, on the other hand, we clamp the boundary with a fixed displacement and allow the stress to fluctuate freely instead. If the constant displacement on the boundary is $\b{u}_{\mathcal{C}}=\Delta_{\mathcal{C}}\hat{\vec{\nu}}$, we have
\begin{equation}
	\oint_{\mathcal{C}}\dd\ell\;\hat{\vec{\nu}}\cdot\b{u}=L_{\mathcal{C}}\Delta_{\mathcal{C}}\;,
\end{equation}
where $L_{\mathcal{C}}$ is the length of the boundary. By using Stokes' theorem, we can rewrite this as a bulk integral,
\begin{equation}
	\dfrac{1}{A}\int\dd^2r\;\vec{\del}\cdot\b{u}=\e\;.\label{eq:e}
\end{equation}
Here we have defined the average strain induced by the boundary as $\e=L_{\mathcal{C}}\Delta_{\mathcal{C}}/A$, where $A=\int\dd\b{r}$ is the area of the sheet. To leading order this strain results in an isotropic change in the projected area as $\delta A/A=\e/2$. As before, $\e>0$ ($\Delta_{\mathcal{C}}>0$) corresponds to a uniform dilational strain, while $\e<0$ ($\Delta_{\mathcal{C}}<0$) is an isotropic compressive strain. We can now integrate out the phonons in either ensemble to get the free energy solely in terms of the flexural modes. In the isotensional ensemble, we obtain
\begin{align}
	\mathcal{F}_{\sigma}&=\int\dd^2r\left[\vphantom{\left(\dfrac{1}{2}\right)^2}\dfrac{\kappa}{2}(\del^2h)^2+\dfrac{\sigma_0}{2}|\vec{\del}h|^2-\mathcal{E}h\right.\nonumber\\
	&\qquad\qquad\left.+\dfrac{Y}{2}\left(\dfrac{1}{2}\mathcal{P}_{ij}^T\partial_ih\partial_jh\right)^2\right]\;.\label{eq:Fsigma}
\end{align}
The subscript $\sigma$ denotes the fixed stress boundary condition imposed and $\mathcal{F}_\sigma$ is the analogous to the Gibbs free energy. The externally applied stress $\sigma_0$ enters in a quadratic term that has been identified previously \cite{guitter1989thermodynamical,kovsmrlj2016response}. A tensile stress ($\sigma_0>0$) suppresses height fluctuations \cite{roldan2011suppression}, while a compressive stress ($\sigma_0<0$) signals the onset of the buckling instability. The Young's modulus controls the now well known nonlinear stretching term \cite{nelson1987fluctuations,bowick2001statistical,nelson2004statistical} via the transverse projection operator ($\mathcal{P}_{ij}^T=\delta_{ij}-\partial_i\partial_j/\del^2$). In the nonlinear stretching term (in Eq.~\ref{eq:Fsigma} and below in Eq.~\ref{eq:Fe}), it is understood that the $\b{q}=\b{0}$ Fourier mode has been projected out \cite{nelson1987fluctuations,bowick2001statistical,nelson2004statistical}, which ensures that $\mathcal{P}^T_{ij}$ is well-defined. This fact also means that the free energy is rotationally invariant when $\sigma_0=0$, i.e., $h(\b{r})\to h(\b{r})+\vec{\alpha}\cdot\b{r}$ where $\alpha_i$ are rotation angles, is a symmetry of the system. The average areal strain ($\e=2\delta A/A$) conjugate to the imposed stress in this ensemble can be computed from the partition function $\mathcal{Z}_{\sigma}=\int\mathcal{D}h\;e^{-\mathcal{F}_{\sigma}/k_BT}$ to give
\begin{equation}
	\langle \e(\sigma_0)\rangle=\dfrac{k_BT}{A}\dfrac{\partial\ln\mathcal{Z}_{\sigma}}{\partial\sigma_0}=\dfrac{\sigma_0}{B}-\dfrac{1}{2A}\int\dd^2r\langle|\vec{\del}h|^2\rangle\;.\label{eq:es}
\end{equation}
The thermal average is computed using $\mathcal{F}_{\sigma}$ and $B=\mu+\lambda$ is the bulk modulus.

In the isometric ensemble, we obtain (see Appendix~\ref{app:iso}) a different result,
\begin{align}
	\mathcal{F}_{\e}&=\int\dd^2r\left\{\dfrac{\kappa}{2}(\del^2h)^2-\mathcal{E}h+\dfrac{Y}{2}\left(\dfrac{1}{2}\mathcal{P}_{ij}^T\partial_ih\partial_jh\right)^2\right.\nonumber\\
	&\qquad\qquad\left.+\dfrac{B}{2}\left[\e+\dfrac{1}{2A}\int\dd^2r|\vec{\del}h|^2\right]^2\right\}\;,\label{eq:Fe}
\end{align}
where the $\e$ subscript now refers to the fixed strain conditions and now $\mathcal{F}_\e$ is analogous to the Helmholtz free energy.
As before we include contributions from bending, the external field and the nonlinear stretching terms. While the Young's modulus penalizes bending induced shear, in the presence of clamped boundaries, dilational stretching induced by flexural deflections can no longer be accommodated by displacing the boundary. Hence, global homogeneous dilations are a zero mode in the isotensional ensemble, but \emph{not} in the isometric ensemble, and are penalized by the bulk modulus in the latter. Upon expanding the final bracket and neglecting an irrelevant constant, we obtain a quadratic term $\sim B\e|\vec{\del}h|^2$ that has been obtained previously \cite{roldan2011suppression} which mirrors the external tension term in Eq.~\ref{eq:Fsigma}. Importantly, we also have an additional \emph{nonlinear} term of the form
\begin{equation}
	\dfrac{B}{8A}\int\dd^2r\int\dd^2r'|\vec{\del}h|^2|\vec{\del}'h|^2\;,\label{eq:term}
\end{equation}
that is independent of the strain imposed, but nonetheless arises only in isometric ensemble. This highly nonlocal term couples the local tilts $\sim\vec{\del}h$ of the membrane at arbitrarily distant points and has been missed in previous studies \cite{guitter1988crumpling,guitter1989thermodynamical,roldan2011suppression,kovsmrlj2016response}. Anisotropic versions of this term do appear in the description of micromechanical resonators as nonlinear beams \cite{lifshitz2008nonlinear} and have been included in a recent mean field analysis of a uniaxially compressed ribbon \cite{hanakata2020thermal}. The consequences of this term in the presence of thermal fluctuations are a major focus of this paper.

We note some further unusual features of the new nonlocal term in Eq.~\ref{eq:term}. Although it involves a double spatial integral, the whole term is extensive (due to the factor of $1/A$), but it is importantly \emph{not} additive. As a result, the membrane cannot be divided into a cumulative sum of macroscopic parts which are roughly independent of each other in the thermodynamic limit. Similar long-ranged interactions appear in models of compressible \cite{sak1974critical,bergman1976critical,de1976coupling} or constrained \cite{rudnick1974renormalization} ferromagnets and can affect critical behaviour in some cases, though without reference to ensemble inequivalence. In self-gravitating systems \cite{padmanabhan1990statistical} and mean-field models of magnets \cite{barre2001inequivalence}, long-ranged interactions are known to spoil the equivalence of canonical and microcanonical ensembles, though typically in the context of first-order phase transitions. Although the buckling instability under compression can proceed as a continuous bifurcation, the highly nonlocal interaction in Eq.~\ref{eq:term} strongly suggests that the isotensional and isometric ensembles may not be equivalent even in the thermodynamic limit ($A\to\infty$).

Before we proceed, we note the isometric ensemble variant of Eq.~\ref{eq:es}. The average stress generated in the sheet due to the imposed strain is simply given by
\begin{equation}
	\langle\sigma(\e)\rangle=-\dfrac{k_BT}{A}\dfrac{\partial\ln\mathcal{Z}_\e}{\partial\e}=B\left(\e+\dfrac{1}{2A}\int\dd^2r\left\langle|\vec{\del}h|^2\right\rangle\right)\;,\label{eq:se}
\end{equation}
with the partition function $\mathcal{Z}_\e=\int\mathcal{D}h\;e^{-\mathcal{F}_\e/k_BT}$ and the thermal average now performed using $\mathcal{F}_\e$. Note that the partition functions in the two ensembles are related,
\begin{equation}
	\mathcal{Z}_\sigma=\mathrm{const}.\;\int_{-\infty}^{\infty}\dd\e\;\mathcal{Z}_\e\;e^{A\sigma_0\e/k_BT}
\end{equation}

In order to incorporate both ensembles within the same calulation, we now work with the following free energy
\begin{align}
	\mathcal{F}&=\int\dd^2r\left[\vphantom{\left(\dfrac{1}{2}\right)^2}\dfrac{\kappa}{2}(\del^2h)^2+\dfrac{\sigma}{2}|\vec{\del}h|^2-\mathcal{E}h\right.\nonumber\\
	&\quad\left.+\dfrac{Y}{2}\left(\dfrac{1}{2}\mathcal{P}_{ij}^T\partial_ih\partial_jh\right)^2+\dfrac{v}{8A}|\vec{\del}h|^2\int\dd^2r'|\vec{\del}'h|^2\right]\;.\label{eq:F}
\end{align}
Upto unimportant additive constants, $\mathcal{F}=\mathcal{F}_\e$ (Eq.~\ref{eq:Fe}) upon identifying $v=B$ and $\sigma=B\e$, where $B=\mu+\lambda$ is the bulk modulus. Alternately, if we set $v=0$ to switch off the nonlocal nonlinear term and set $\sigma=\sigma_0$, then we find $\mathcal{F}=\mathcal{F}_\sigma$ (Eq.~\ref{eq:Fsigma}). It is important to note that setting $v=0$ is only a mathematical trick to obtain $\mathcal{F}_\sigma$ from Eq.~\ref{eq:F}. It does \emph{not} imply that the actual bulk modulus vanishes. As we shall see later, the nonlocal nature of this new term guarantees that it cannot be generated if initially absent, i.e., if we set $v=0$ in Eq.~\ref{eq:F}, then it remains zero under the coarse-graining embodied in a renormalization group transformation. Additionally we will see that the actual physical elastic moduli ($\mu,\lambda$) renormalize identically in both ensembles, but the buckling transition is described by two distinct fixed points (with distinct critical exponents for some quantities!) depending on the ensemble. So, in all that follows, we will use $v$ as a coupling constant that distinguishes the two ensembles with $v=0$ being allowed in the isotensional ensemble and $v>0$ only being allowed in the isometric ensemble. Only in the latter case will $v$ be identical to the actual bulk modulus ($B$) of the sheet.

\section{Definition of scaling exponents near buckling}
\label{sec:exp}
Before we analyze buckling criticality, we define our notation. Close to the buckling transition, we expect power law scaling in a number of quantities, the exponents for which we define below. The unprimed exponents below will refer to the isotensional (constant stress) ensemble and the primed exponents to the isometric (constant strain) ensemble. We will denote the buckling threshold as $\sigma_c$ and $\e_c$ in the isotensional and isometric ensembles respectively.

\subsection{Mechanical response}
Upon approaching the buckling transition, the sheet develops a variety of anomalous mechanical responses. In the absence of an external field ($\mathcal{E}=0$), up/down symmetry is spontaneously broken and the sheet develops a finite $\langle h\rangle\neq 0$ when buckled. The average height rises continuously at the transition, acting as an order parameter,
\begin{equation}
	\langle h\rangle\propto\begin{cases}
		|\sigma_0-\sigma_c|^\beta\quad (\textrm{isotensional})\;,\\
		|\e-\e_c|^{\beta'}\quad (\textrm{isometric})\;.
	\end{cases}\label{eq:beta}
\end{equation}
The zero field susceptibility exhibits divergent scaling near buckling,
\begin{equation}
	\chi=\left.\dfrac{\partial\langle h\rangle}{\partial\mathcal{E}}\right|_{\mathcal{E}=0}\propto\begin{cases}
		|\sigma_0-\sigma_c|^{-\gamma}\quad (\textrm{isotensional})\;,\\
		|\e-\e_c|^{-\gamma'}\quad (\textrm{isometric})\;.
	\end{cases}\label{eq:gamma}
\end{equation}
We expect the exponents $\gamma,\gamma'$ to be the same on either side of the transition \cite{goldenfeld2018lectures}, though the amplitudes of the scaling function can and will be different. The divergence of the susceptibility signals the breakdown of linear response, which is also seen in the nonlinear field dependence right at the buckling transition
\begin{equation}
	\langle h\rangle\propto\begin{cases}
		\mathcal{E}^{1/\delta}\quad (\sigma_0=\sigma_c\;,\ \textrm{isotensional})\;,\\
		\mathcal{E}^{1/\delta'}\quad (\e=\e_c\;,\,\textrm{isometric})\;.
	\end{cases}\label{eq:delta}
\end{equation}
Finally, in conjunction with these out-of-plane responses, we also have a concomitant nonlinear scaling in the in-plane mechanics, quantified by anomalous stress-strain curves,
\begin{subequations}
\begin{align}
	\langle\e\rangle&\propto\rm{const.}+(\sigma_0-\sigma_c)^{1/\theta}\quad(\textrm{isotensional})\;,\\
	\langle\sigma\rangle&\propto\rm{const.}+(\e-\e_c)^{\theta'}\quad(\textrm{isometric})\;,
\end{align}\label{eq:theta}
\end{subequations}
at zero external field ($\mathcal{E}=0$). Note that the above only includes the dominant singularity and neglects other contributions. The new exponents $\theta,\theta'\neq 1$ signal a violation of Hooke's law.

\subsection{Fluctuations and spatial scales}
Apart from global quantites discussed above, local variables also develop extended correlations when near the buckling transition. In the absence of an external field ($\mathcal{E}=0$), the fluctuating height of the sheet has spatial correlations with nontrival scaling properties. A nonzero external stress or strain generically causes the height fluctuations to decay exponentially over a finite correlation length $\xi$, albeit with different large distance asymptotes depending on whether the sheet is buckled or flat. This behaviour is also reflected in the normal-normal correlation function. The unit normal to a surface specified by $\b{X}(\b{r})=(x,y,h(\b{r}))$ in the Monge representation, is $\hat{\b{n}}=(-\partial_xh,-\partial_yh,1)/\sqrt{1+|\vec{\del}h|^2}$, which allows us to simply relate the normal-normal and height-height correlation functions as
\begin{equation}
	\langle\hat{\b{n}}(\b{r})\cdot\hat{\b{n}}(\b{0})\rangle\simeq1-\dfrac{1}{2}\left\langle\left|\vec{\del}h(\b{r})-\vec{\del}h(\b{0})\right|^2\right\rangle\;,
\end{equation}
at lowest order in the height gradients. On either side of the buckling transition, we have $\langle\hat{\b{n}}(\b{r})\cdot\hat{\b{n}}(\b{0})\rangle\sim e^{-r/\xi}$, neglecting asymptotic constants and nonexponential prefactors. The correlation length diverges at buckling as
\begin{equation}
	\xi\propto\begin{cases}
		|\sigma_0-\sigma_c|^{-\nu}\quad (\textrm{isotensional})\;,\\
		|\e-\e_c|^{-\nu'}\quad (\textrm{isometric})\;.
	\end{cases}\label{eq:nu}
\end{equation}
Right at the buckling transition, the normal correlations decay as a power law and the sheet has critical fluctuations on all scales that cause the correlation functions to behave anomalously. We define the translationally invariant height and phonon correlators well away from the boundaries as
\begin{align}
	G_h(\b{r})&=\langle h(\b{r})h(\b{0})\rangle\;,\\
	\left[\b{G}_u(\b{r})\right]_{ij}&=\langle u_i(\b{r})u_j(\b{0})\rangle\;.
\end{align}
Upon tuning to the transition, the Fourier transformed correlators [$G(\b{q})=\int\dd\b{r}\;e^{-i\b{q}\cdot\b{r}}G(\b{r})$] exhibit a power law scaling as $q=|\b{q}|\to0$. These averages define the well known anomalous exponents $\eta$ and $\eta_u$ \cite{nelson1987fluctuations,le1992self,*le2018anomalous,david1988crumpling,aronovitz1988fluctuations,*aronovitz1989fluctuations,bowick1996flat} through
\begin{equation}
	G_h(\b{q})\propto\begin{cases}
		q^{-(4-\eta)}\quad (\sigma_0=\sigma_c\;,\,\textrm{isotensional})\;,\\
		q^{-(4-\eta')}\quad (\e=\e_c\;,\,\textrm{isometric})\;,
	\end{cases}\label{eq:eta}
\end{equation}
and for the in-plane phonons (irrespective of the tensor indices),
\begin{equation}
	\b{G}_u(\b{q})\propto\begin{cases}
		q^{-(2+\eta_u)}\quad (\sigma_0=\sigma_c\;,\,\textrm{isotensional})\;,\\
		q^{-(2+\eta'_u)}\quad (\e=\e_c\;,\,\textrm{isometric})\;.
	\end{cases}\label{eq:etau}
\end{equation}
These results describe a divergent renormalization of the wave-vector dependent bending rigidity $\kappa(\b{q})\sim q^{-\eta}$ and softening of the elastic moduli $\mu(\b{q}),\lambda(\b{q})\sim q^{\eta_u}$ (with analogous expressions with exponents $\eta'$ and $\eta'_u$ in the isometric ensemble).
\section{Mean field theory}
\label{sec:mft}
\begin{figure}[]
	\centering{\includegraphics[width=0.3\textwidth]{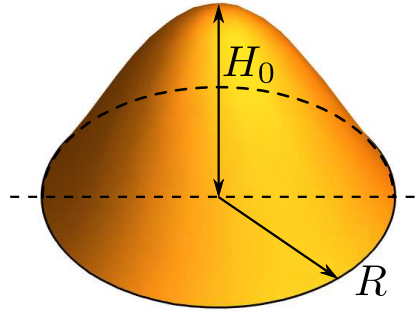}}
	\caption{Sketch of the first buckling mode with boundary conditions such that $h_0(r=R)=0$, for a circular plate of radius $R$. The amplitude of the mode at the center of the circular frame is $H_0$ and its wavevector is $q_0\sim1/R$. We assume hinged boundary conditions at $r=R$ for simplicity. Qualitatively similar buckling modes appear for alternative boundary conditions, such as for membranes that approach $r=R$ tangentially.}
	\label{fig:buckled}
\end{figure}

Here we neglect thermal fluctuations and analyze the buckling transition in the mean-field limit, as appropriate at $T=0$. By minimizing the free energy $\mathcal{F}$ (Eq.~\ref{eq:F}) over the surface profile $h(\b{r})$, we obtain the Euler-Lagrange equation,
\begin{align}
	&\kappa\del^4h-\sigma\del^2h-Y\left[\mathcal{P}_{k\ell}^T\left(\dfrac{1}{2}\mathcal{P}_{ij}^T\partial_ih\partial_jh\right)\right]\partial_k\partial_\ell h\nonumber\\
	&\quad-\dfrac{v}{2A}\del^2h\int\dd^2r'|\vec{\del}'h|^2=\mathcal{E}\;,\label{eq:EL}
\end{align}
where $v=0$ and $v=B>0$ again allows us to distinguish ensembles.
As both the elastic sheet and the external load are isotropic, axisymmetry is assumed in the following. We choose the eigenfunction of the linearized operator in Eq.~\ref{eq:EL} as an ansatz for the buckled height profile in a circular geometry,
\begin{equation}
	h_0(\b{r})=H_0J_0(q_nr)\;.
\end{equation}
Here, $r=|\b{r}|$ is the distance from the center of the disc and $J_0(x)$ is the Bessel function of the first kind. The mode of buckling is controlled by $q_n$ which is fixed by boundary conditions on the height. For simplicity, we shall assume
\begin{equation}
	h_0(R)=0\implies J_0(q_nR)=0\;,\label{eq:bc}
\end{equation}
where $R$ is the radius of the sheet. Other boundary conditions can also be easily used with only minor quantitative changes in the results. A simple Galerkin approximation \cite{galerkin1915series} involves projecting Eq.~\ref{eq:EL} onto the single mode ansatz, which then gives (the details of the calculation are provided in Appendix~\ref{app:mft})
\begin{equation}
	q_n^2(\kappa q_n^2+\sigma)H_0+q_n^4\left[\dfrac{v}{2}f(q_nR)+c_0Y\right]H_0^3=\mathcal{E}q_nR\;,\label{eq:eos}
\end{equation}
where $c_0\approx0.10567$ is a constant and $f(x)$ is a dimensionless function given in Appendix~\ref{app:mft}, with the asymptotics $f(x)\sim2/(\pi x)$ for $x\to\infty$. Similar to the simpler problem of the buckling of a ribbon at $T=0$ \cite{hanakata2020thermal}, Eq.~\ref{eq:eos} resembles a Landau theory with $H_0$ as the order parameter. It is the mechanical equivalent of a mean field ``equation of state''. Note, however, that the underlying Landau theory has coefficients that depend on system size. In the absence of an external field ($\mathcal{E}=0$), buckling occurs for a sufficiently negative $\sigma$ and spontaneously breaks up-down inversion symmetry. At the buckling threshold, the lowest ($n=0$) mode (shown in Fig.~\ref{fig:buckled}) goes unstable first. The buckling amplitude in either ensemble is given by
\begin{equation}
	|H_0|=\begin{cases}
		\left(\dfrac{\sigma_{c}-\sigma_0}{c_0Yq_0^2}\right)^{1/2}\quad(\sigma_0<\sigma_c\;,\,\textrm{isotensional})\;,\\
		\left(\dfrac{\e_{c}-\e}{c'_0q_0^2}\right)^{1/2}\quad(\e<\e_c\;,\,\textrm{isometric})\;,
	\end{cases}\label{eq:H0}
\end{equation}
where $c'_0=f(m_0)+2c_0(Y/B)$ is weakly dependent on the Poisson's ratio through $Y/B$. As expected, we obtain a standard square root scaling typical of pitchfork bifurcations ($\beta=\beta'=1/2$).
For both $\sigma_0>\sigma_{c}$ or $\e>\e_{c}$, we have a stable flat state with $H_0=0$. The critical stress for buckling is $\sigma_c=-\kappa q_0^2$ and the critical strain is $\e_c=-\kappa q_0^2/B$ in the respective ensembles. The wavevector that first goes unstable with decreasing $\sigma_0<0$ or $\e<0$ is $q_0=m_0/R$, where $m_0\approx2.405$ is the smallest positive root of $J_0(m_0)=0$ as required by the boundary condition (Eq.~\ref{eq:bc}). At threshold, $f(m_0)=J_1(m_0)^2\approx0.27$, is finite.

A couple of points are worth remarking on here. The buckling thresholds in both ensembles involve compression ($\sigma_c,\e_c<0$) and are $\propto 1/R^2$, vanishing as the area of the sheet becomes larger. Thus, in the thermodynamic limit ($R\to\infty$), classical buckling is a thresholdless long-wavelength ($q_0\sim1/R\to 0$) instability. At the same time, the buckling amplitude remains macroscopic ($|H_0|\propto R$). Note also that, for a circular geometry, the buckled state acquires nonzero Gaussian curvature due to the isotropic nature of the loading. This is the energetically preferred state: a uniaxially buckled sheet that remains developable has a higher energy for these circular loading conditions. As a result, even in the isotensional ensemble, we pay stretching energy $\sim Y$ upon buckling (the penalty associated with Gaussian curvature in Eq.~\ref{eq:F}), while in the isometric ensemble, both the bulk ($B$) and Young's ($Y$) moduli contribute.

Right at the transition, in the presence of an external field ($\mathcal{E}\neq 0$), we have an nonlinear response of height given by
\begin{equation}
	H_0=\begin{cases}
		\left(\dfrac{m_0\mathcal{E}}{c_0Yq_0^4}\right)^{1/3}\quad(\sigma_0=\sigma_c\;,\,\textrm{isotensional})\;,\\
		\left(\dfrac{2m_0\mathcal{E}}{c'_0Bq_0^4}\right)^{1/3}\quad(\e=\e_c\;,\,\textrm{isometric})\;.
	\end{cases}\label{eq:H0E}
\end{equation}
As is typical of mean field models, we obtain $\delta=\delta'=3$.
Note that, because $q_0\sim 1/R$, this response is strongly size dependent, diverging as $R^{4/3}$ in a large sheet. Finally, we also have the zero field susceptibility
\begin{equation}
	\chi=\left.\dfrac{\partial H_0}{\partial\mathcal{E}}\right|_{\mathcal{E}=0}=\begin{cases}
		c_{\pm}\dfrac{m_0}{q_0^2|\sigma_0-\sigma_c|}\quad(\textrm{isotensional})\;,\\
		c_{\pm}\dfrac{m_0}{q_0^2B|\e-\e_c|}\quad(\textrm{isometric})\;,
	\end{cases}\label{eq:chi}
\end{equation}
which diverges right at the transition with $\gamma=\gamma'=1$. The magnitude of this divergence is different on either side of the transition, with $c_+=1$ above the transition and $c_-=1/2$ below the transition, irrespective of the ensemble. Once again we find a strong size dependence, with $\chi\propto R^2$ diverging as the area.
Finally, a simple calculation using Eqs.~\ref{eq:es} and \ref{eq:se} also determines the stress-strain relation to be $\sigma_0\propto\e$, which sets the exponents defined by Eq.~\ref{eq:theta} to be $\theta=\theta'=1$ in both ensembles.
Unlike the previous expressions, the stress-strain relation is system size independent. Although the above analysis was restricted to 2D membranes deforming in 3D space, all of the mean-field results are qualitatively similar in arbitrary dimensions.
%\begin{subequations}
%\begin{align}
%	&\textrm{Isotensional:}\nonumber\\
%	&\langle\e\rangle=\begin{cases}
%		\dfrac{\sigma_0}{B}\quad(\sigma_0>\sigma_c)\;,\\
%		\sigma_0\dfrac{c_0'}{2c_0Y}-\dfrac{f(m_0)\sigma_c}{2c_0Y}\quad(\sigma_0<\sigma_c)\;,
%	\end{cases}\\
%	&\textrm{Isometric:}\nonumber\\
%	&\langle\sigma\rangle=\begin{cases}
%		B\e\quad(\e>\e_c)\;,\\
%		\left(\dfrac{B}{2}+\dfrac{c_0Y}{c'_0}\right)\e+\dfrac{Bf(m_0)\e_c}{2c'_0}\quad(\e<\e_c)\;,
%	\end{cases}
%\end{align}
%\end{subequations}

It is evident that the two ensembles have equivalent scaling behaviour in the mean-field limit. In addition, some of the scaling functions have a nontrivial size dependence, a feature peculiar to the buckling transition. This size dependence is unusual from the point of view of conventional critical phenomena \cite{goldenfeld2018lectures}, and as we shall see in Sec.~\ref{sec:scaling}, it has important consequences for the critical exponents and their scaling relations. Below we go beyond the mean field limit by including thermal fluctuations and show that there are important changes.

\section{Gaussian analysis}
\label{sec:gaussian}
Here we shall primarily consider the simpler case of a flat unbuckled membrane with $H_0=0$ with a vanishing symmetry-breaking external field ($\mathcal{E}=0$). We can rewrite Eq.~\ref{eq:F} as $\mathcal{F}=\mathcal{F}_0+\mathcal{F}_{\rm int}$ where
\begin{align}
	\mathcal{F}_0&=\dfrac{1}{2}\int\dd^2r\left[\kappa(\del^2h)^2+\sigma|\vec{\del}h|^2\right]\;,\\
	\mathcal{F}_{\rm int}&=\dfrac{Y}{2}\int\dd^2r\left(\dfrac{1}{2}\mathcal{P}_{ij}^T\partial_ih\partial_jh\right)^2\nonumber\\
	&\quad+\dfrac{v}{8A}\int\dd^2r\int\dd^2r'|\vec{\del}h|^2|\vec{\del}'h|^2\;.\label{eq:Fint}
\end{align}
For small fluctuations, one might hope to neglect the nonlinear terms in $\mathcal{F}_{\rm int}$. Upon Fourier transforming ($h_{\b{q}}\equiv\int\dd\b{r}\;e^{-i\b{q}\cdot\b{r}}h(\b{r})$), we obtain the bare height-height correlation function
\begin{equation}
	G^0_h(\b{q})=\dfrac{1}{A}\langle |h_{\b{q}}|^2\rangle_0=\dfrac{k_BT}{\kappa q^4+\sigma q^2}\;,\label{eq:g0h}
\end{equation}
where $q=|\b{q}|$ and the zero subscript denotes that the average is performed in the noninteracting limit. We can similarly neglect all the nonlinear interactions in $\mathcal{H}$ (Eq.~\ref{eq:H}) to find the bare in-plane phonon correlation function,
\begin{align}
	\left[\b{G}^0_u(\b{q})\right]_{ij}&=\dfrac{1}{A}\langle u_i(\b{q})u_j(-\b{q})\rangle_0\nonumber\\
	&=\dfrac{k_BT}{\mu q^2}\mathcal{P}_{ij}^T(\b{q})+\dfrac{k_BT}{(2\mu+\lambda)q^2}\mathcal{P}_{ij}^L(\b{q})\;,
\end{align}
involving the longitudinal ($\mathcal{P}_{ij}^L(\b{q})=q_iq_j/q^2$) and the transverse ($\mathcal{P}_{ij}^T(\b{q})=\delta_{ij}-q_iq_j/q^2$) projection operators.
As is evident, at the Gaussian level, we have $\eta=\eta'=0$ and $\eta_u=\eta'_u=0$ in both ensembles. From Eq.~\ref{eq:g0h}, we can easily show that the correlation length in Gaussian limit is
\begin{equation}
	\xi=\sqrt{\dfrac{\kappa}{|\sigma|}}\propto|\sigma|^{-1/2}\;,
\end{equation}
which corresponds to $\nu=\nu'=1/2$ in both ensembles. Note that, here we work in the large sheet limit, which allows us to freely Fourier transform and set $\sigma_c,\e_c\sim 0$.

We can now determine the importance of the nonlinear terms in Eq.~\ref{eq:Fint} by making a scale transformation. To do this, we rescale $\b{r}\to b\b{r}$, $h\to b^{\zeta}h$ to get the following scaling dimensions for the bending rigidity, tension and nonlinear couplings,
\begin{equation}
	y_{\kappa}=2\zeta-2\;,\quad y_{\sigma}=2\zeta\;,\quad y_Y=y_v=4\zeta-2\;.
\end{equation}
where we have used the fact that the area $A\to b^2A$ under scaling. Here, $D=2$ and $d=3$, but these scalings depend more generally on dimensionality;their generalization for general $D$-dimensional manifolds embedded in $d$-dimensions is given in Appendix~\ref{app:rgd}.
In the Gaussian limit, we have $\zeta=1$, as $h$ is simply the height with na{\"i}ve dimensions of length. This leaves the bending term scale-invariant ($y_{\kappa}=0$), but the external load ($\sigma$), Young's modulus ($Y$) and nonlocal coupling ($v$) are all equally relevant perturbations for 2D membranes embedded in 3D: $y_\sigma=y_Y=y_v=2>0$.

Hence, even at low temperatures when fluctuations may be small, we expect that the nonlinear interactions eventually dominate in a large enough sheet. As usual, a Ginzburg-like criterion determines the thermal length scale beyond which such nonlinear fluctuations dominate \cite{nelson1987fluctuations,kantor1987phase,kovsmrlj2016response,bowick2017non}
\begin{equation}
	\ell_{\rm{th}}=\sqrt{\dfrac{16\pi^3\kappa^2}{3k_BTY}}\;.\label{eq:lth}
\end{equation}
Remarkably, at room temeprature, a monolayer of graphene or MoS$_2$ has $\ell_{\rm{th}}\sim 1-10~\rm{\AA}$, and thermal fluctuations matter already on the atomic scale. Softer materials, such as naturally occuring 2D organic polymers \cite{schmidt1993existence,hermanson2007engineered,klotz2020equilibrium} have a typical $\ell_{\rm{th}}\sim0.1-1~\mu$m range which is much larger due to their smaller Young's moduli. As a result, the consequences of thermal fluctuations are most dramatic in atomic crystals in contrast to the other soft membranes. We can perturbatively account for such fluctuation effects within a renormalization group framework that we implement below.

\section{Perturbative Renormalization Group}
\label{sec:rg}
We now implement a conventional Wilsonian renormalization group \cite{goldenfeld2018lectures} by iteratively integrating out a thin shell in momentum space of short wavelength fluctuations. The cutoff in Fourier space is $\Lambda\sim1/a$, where $a$ is the microscopic lattice spacing. As an aside, we note that, although the nonlocal term involving $v$ in Eq.~\ref{eq:F} is quite unusual, it can be treated straightforwardly within a standard Wilsonian treatment, as has been done for related problems in, for example, compressible magnets \cite{sak1974critical,bergman1976critical,de1976coupling}. We perform a systematic $\vare=4-D$ expansion about the upper critical dimension following previous works on unconstrained sheets \cite{aronovitz1988fluctuations,guitter1988crumpling}. Although the full diagrammatic calculation is presented in Appendix~\ref{app:rgd}, we describe the main results below. In Appendix~\ref{app:rg}, we separately provide a simple, but uncontrolled, one-loop calculation with fixed internal ($D=2$) and external ($d=3$) dimensions that is qualitatively correct, but numerically inaccurate.
%The latter will prove useful later to perform a systematic $D=4-\vare$ calculation, whose results we quote directly in Sec.~\ref{sec:scaling} and Table~\ref{tab:exponents}.
\subsection{Recursion relations}
We carry out a perturbative low temperature expansion evaluation of thermal fluctuations to one loop order. By integrating out fluctuations within a shell of wavevectors $\Lambda/b\leq q\leq\Lambda$, where $b=e^{s}$ is a scale factor, and $\Lambda^{-1}$ a short distance cut-off of order the lattice spacing or membrane thickness, we compute corrections to the various parameters in the model. As explained in Appendix~\ref{app:rgd}, even with the addition of the new nonlinear term, the form of our elastic description in $\mathcal{F}$ (Eq.~\ref{eq:F}) remains unchanged at long-wavelengths under coarse-graining; only coupling constants such as $\kappa,\sigma,Y,v$ and $\mathcal{E}$ get renormalized. The fluctuation corrections can be cast as differential recursion relations given below
\begin{align}
	\dfrac{\dd\kappa}{\dd s}&=\kappa(2\zeta-\vare)+\dfrac{5k_BT\Lambda^2}{192\pi^2(\kappa\Lambda^2+\sigma)}(Y+4\mu)\;,\label{eq:dkappadl}\\
	\dfrac{\dd\sigma}{\dd s}&=\sigma(2\zeta+2-\vare)+\dfrac{k_BTv\Lambda^4}{16\pi^2(\kappa\Lambda^2+\sigma)}\;,\label{eq:dsigmadl}\\
	\dfrac{\dd Y}{\dd s}&=Y(4\zeta-\vare)-\dfrac{5k_BTY^2\Lambda^4}{384\pi^2(\kappa\Lambda^2+\sigma)^2}\;,\label{eq:dYdl}\\
	\dfrac{\dd \mu}{\dd s}&=\mu(4\zeta-\vare)-\dfrac{k_BT\mu^2\Lambda^4}{96\pi^2(\kappa\Lambda^2+\sigma)^2}\;,\label{eq:dmudl}\\
	\dfrac{\dd v}{\dd s}&=v(4\zeta-\vare)-\dfrac{k_BTv^2\Lambda^4}{16\pi^2(\kappa\Lambda^2+\sigma)^2}\;,\label{eq:dvdl}\\
	\dfrac{\dd\mathcal{E}}{\dd s}&=\mathcal{E}(4-\vare+\zeta)\;.\label{eq:dEdl}
\end{align}
The fluctuation corrections are evaluated here to leading order in $\vare=4-D$ and the codimension of the manifold is set to its physically relevant value of $d_c=d-D=1$. Furthermore, as these equations are derived in general dimension, we use the $D$-dimensional generalization of the Young's modulus ($Y=2\mu(2\mu+D\lambda)/(2\mu+\lambda)$) and the bulk modulus ($B=(2\mu/D)+\lambda$), which reduce to the standard 2D expressions for $D=2$, as expected.

The renormalization equations for $\kappa$, $\mu$ and $Y$ in Eqs.~\ref{eq:dkappadl},~\ref{eq:dmudl} and~\ref{eq:dYdl} are identical to those obtained previously \cite{aronovitz1988fluctuations,guitter1989thermodynamical}, while the important coupled equations for $\sigma$ and $v$ are new results. The difference between the isometric and isotensional ensembles captured by the presence of the nonlinear coupling $v$ is already reflected in the modified renormalization group flows. We have also retained the external field $\mathcal{E}$ here \footnote{As we have set $d_c=1$ here, both the height $h$ and the symmetry breaking field $\mathcal{E}$ are scalars here.}; this quantity renormalizes trivially without any graphical corrections as it couples only to the average height ($\int\dd\b{r}\;h=h_{\b{q}=\b{0}}$), which is left untouched by the elastic and geometric nonlinearities.

The shear and Young's moduli renormalize independently, as expected, but they both contribute to the renormalization of the bending rigidity near $D=4$. One can easily use the recursion relations for $\mu$ and $Y$ to obtain equivalent ones for the $D$-dimensional versions of the bulk modulus ($B=(2\mu/D)+\lambda$) and the Poisson's ratio ($\nu_p=\lambda/[2\mu+(D-1)\lambda]$), namely
\begin{align}
	\dfrac{\dd B}{\dd s}&=B(4\zeta-\vare)-\dfrac{k_BTB^2\Lambda^4}{16\pi^2(\kappa\Lambda^2+\sigma)^2}\;,\label{eq:dBdl}\\
	\dfrac{\dd\nu_p}{\dd s}&=-\dfrac{k_BT\mu\Lambda^4}{192\pi^2(\kappa\Lambda^2+\sigma)^2}(1+\nu_p)(1+3\nu_p)\;.\label{eq:dnudl}
\end{align}

%Although, only the Young's modulus enters $\mathcal{F}$ within this reduced description of 2D sheets, we can independently compute the fluctuation corrections to the shear ($\mu$) and bulk ($B=\mu+\lambda$) using the original full elastic Hamiltonian $\mathcal{H}$. At one loop order (the details are given in Appendix~\ref{app:rg}), we find
%\begin{align}
%	\dfrac{\dd\mu}{\dd s}&=\mu(4\zeta-2)-k_BT\dfrac{\mu^2\Lambda^2}{8\pi(\kappa\Lambda^2+\sigma)^2}\;,\label{eq:dmudl}\\
%	\dfrac{\dd B}{\dd s}&=B(4\zeta-2)-k_BT\dfrac{B^2\Lambda^2}{4\pi(\kappa\Lambda^2+\sigma)^2}\;,\label{eq:dBdl}\\
%	\dfrac{\dd\nu_p}{\dd s}&=-k_BT\dfrac{\mu\Lambda^2}{16\pi(\kappa\Lambda^2+\sigma)^2}(1+\nu_p)(1+3\nu_p)\;.\label{eq:dnudl}
%\end{align}
%We have additionally included the flow equation for the 2D Poisson's ratio $\nu_p=\lambda/(2\mu+\lambda)$ for completeness sake, though it is fully determined by Eqs.~\ref{eq:dmudl} and~\ref{eq:dBdl}.
A couple of points are worth noting here. First, as expected, the bulk and shear moduli also renormalize independently. Second, upon comparing Eq.~\ref{eq:dBdl} and Eq.~\ref{eq:dvdl}, we immediately see that both $v$ and $B$ renormalize in identical ways, guaranteeing that in the isometric ensemble, since $v=B$ at the microscopic scale, they remain equal on larger scales as well. In contrast, the isotensional ensemble is characterized by $v=0$ (which remains invariant under renormalization), even though $B>0$.

The third important point concerns the Poisson's ratio $\nu_p$ \footnote{Note that, we define the Poisson's ratio simply using the elastic moduli here, as is usually done \cite{le2018anomalous,nelson2004statistical}. Alternate, yet related, definitions are possible that can yield different results, see Ref.~\cite{burmistrov2018differential} for instance.}. As is easily seen from Eq.~\ref{eq:dnudl}, we have a stable fixed point where $\dd\nu_p/\dd s=0$ at $\nu_p=-1/3$ ($\nu_p=-1$ is unphysical as it corresponds to $\lambda=-2\mu/D$ leading to a marginally stable solid with vanishing bulk and Young's moduli), which is exactly the one-loop estimate for the universal Poisson's ratio of a free standing elastic membrane, in accord with previous self consistent calculations \cite{le1992self,*le2018anomalous} and Monte-Carlo simulations \cite{bowick1996flat,falcioni1997poisson,*bowick2001universal,cuerno2016universal}. This universal auxetic response is a characteristic property of the flat phase of unconstrained thermalized membranes \cite{bowick2001universal,bowick2001statistical}. The reason we obtain this result from a simple one-loop calculation near $D=4$ is that the structure of the one-loop calculation is the same as that of the self-consistent calculation done by \citet{le1992self}, and the higher order box diagrams are convergent in that case. Furthermore, the universal Poisson's ratio obtained is independent of both internal and embedding dimensions of the membrane \cite{le1992self}, hence we recover the same value even in an $\vare=4-D$ expansion.

To analyze these recursion relations, we introduce the following dimensionless variables (the Poisson's ratio is of course already dimensionless),
\begin{gather}
	K=\dfrac{\kappa\Lambda^2}{\kappa\Lambda^2+\sigma}\;,\quad\bar{Y}=\dfrac{k_BT\Lambda^4}{(\kappa\Lambda^2+\sigma)^2}Y\;,\nonumber\\
	\bar{\mu}=\dfrac{k_BT\Lambda^4}{(\kappa\Lambda^2+\sigma)^2}\mu\;,\quad\bar{v}=\dfrac{k_BT\Lambda^4}{(\kappa\Lambda^2+\sigma)^2}v\;,
\end{gather}
which are appropriate near $D=4$ dimensions. For general $D$, we must replace the factor of $\Lambda^4$ by $\Lambda^D$ to keep $\bar{Y}$, $\bar{\mu}$ and $\bar{v}$ dimensionless.
As the external field $\mathcal{E}$ does not influence the fixed points, we will set $\mathcal{E}=0$ for now. The effects of $\mathcal{E}\neq 0$ will be addressed within a general scaling theory we develop in Sec.~\ref{sec:scaling}.

The physical interpretation of $K$ is that it measures the relative importance of bending to the external load. We note that $0\leq K<\infty$ and then demarcate three distinct regimes based on the value for $K$ as follows:
\begin{itemize}
	\item $0<K<1$ : Tension dominated ($\sigma>0$),
	\item $K\sim 1$ : Bending dominated ($\sigma\sim 0$),
	\item $K>1$ : Compression dominated ($\sigma<0$).
\end{itemize}
The buckled state we found using mean-field theory thus occurs for $K>1$ in the presence of compression.
The recursion relations for these dimensionless coupling constants then read
\begin{align}
	\dfrac{\dd K}{\dd s}&=2(K-1)\left[K-\dfrac{5}{384\pi^2}(\bar{Y}+4\bar{\mu})\right]-\dfrac{\bar{v}K}{16\pi^2}\;,\label{eq:Krg}\\
	\dfrac{\dd\bar{Y}}{\dd s}&=\left[\vare+4(K-1)-\dfrac{\bar{v}}{8\pi^2}-\dfrac{25\bar{Y}}{384\pi^2}-\dfrac{5\bar{\mu}}{24\pi^2}\right]\bar{Y}\;,\label{eq:Yrg}\\
	\dfrac{\dd\bar{\mu}}{\dd s}&=\left[\vare+4(K-1)-\dfrac{\bar{v}}{8\pi^2}-\dfrac{5\bar{Y}}{96\pi^2}-\dfrac{7\bar{\mu}}{32\pi^2}\right]\bar{\mu}\;.\label{eq:murg}\\
	\dfrac{\dd\bar{v}}{\dd s}&=\left[\vare+4(K-1)-\dfrac{3\bar{v}}{16\pi^2}-\dfrac{5\bar{Y}}{96\pi^2}-\dfrac{5\bar{\mu}}{24\pi^2}\right]\bar{v}\;.\label{eq:vrg}\\
	\dfrac{\dd\nu_p}{\dd s}&=-\dfrac{k_BT\mu\Lambda^4}{192\pi^2(\kappa\Lambda^2+\sigma)^2}(1+\nu_p)(1+3\nu_p)\;.\label{eq:nurg}
\end{align}
As expected, the scale factor $\zeta$ for the flexural phonon field drops out of the equations when cast in terms of dimensionless variables.
Note that, while we quote the renormalization group flow equations for $\bar{Y},\bar{\mu}$ and $\nu_p$ (Eqs.~\ref{eq:Yrg},~\ref{eq:murg} and~\ref{eq:nurg}), only two of the three are independent, as $\nu_p=(Y-2\mu)/[2\mu+(D-2)Y]$.
For an unconstrained membrane, $K$ remains fixed at unity, as $\sigma=\bar{v}=0$. While in the isotensional ensemble, we can have any $\sigma\neq 0$ ($K\neq 1$) with $\bar{v}=0$, in the isometric ensemble, we generally have \emph{both} $\sigma\neq 0$ and $\bar{v}\neq 0$. In the latter case, the system spontaneously develops a thermally generated tension due to the geometric confinement enforced by the clamped boundary conditions, as discussed below. But first, we analyze the fixed points of the recursion relations.
\subsection{Fixed points}
\begin{figure*}[]
	\centering{\includegraphics[width=0.8\textwidth]{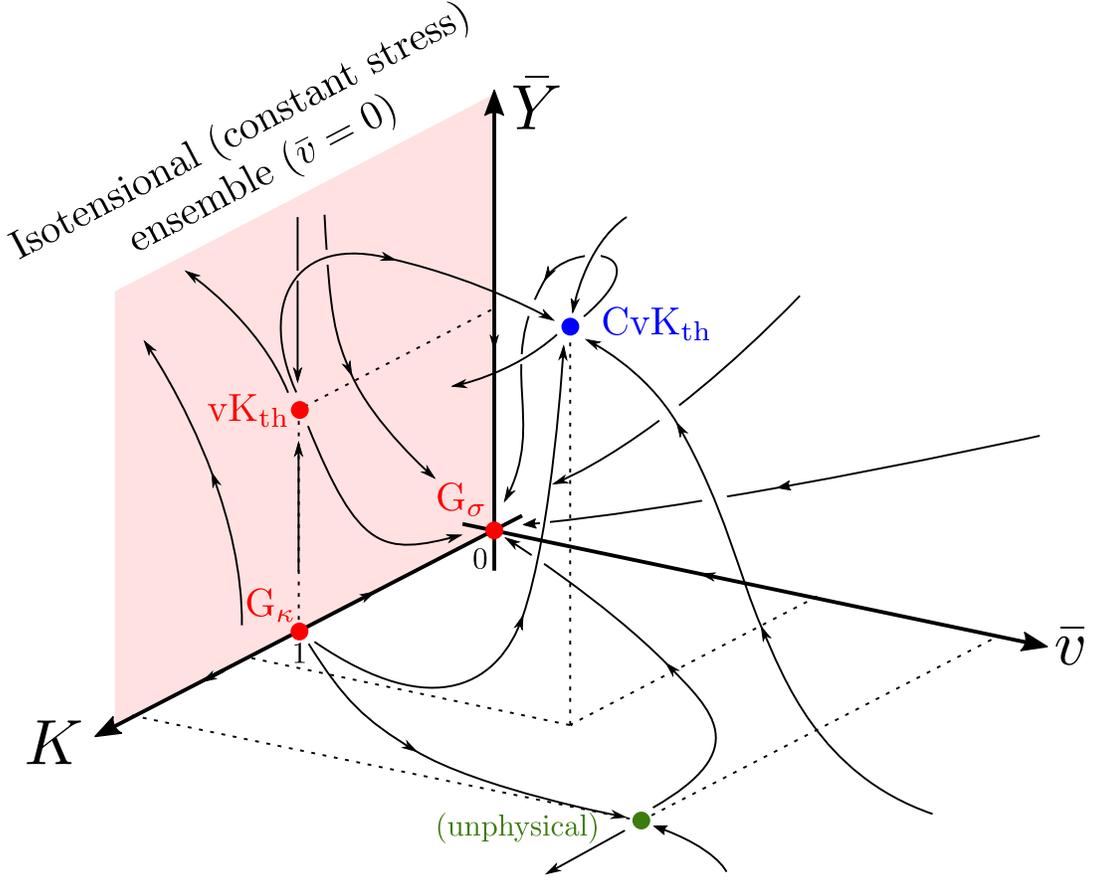}}
	\caption{A schematic of the full renormalization group flow diagram in the three-dimensional parameter space of $\{K,\bar{Y},\bar{v}\}$. We fix the Poisson's ratio to its universal value $\nu_p=-1/3$ here, so that both the shear and bulk moduli are determined by the Young's modulus through $\bar{\mu}=(D+1)\bar{Y}/4$ and $\bar{B}=(D+1)\bar{Y}/D(D+2)$. The three isotensional fixed points (G$_\sigma$, G$_\kappa$ and vK$_{\rm{th}}$) are shown as red points with $0\leq K\leq 1$ and $\bar{v}=0$, while the new constrained fixed point CvK$_{\rm th}$ with $K>1$ and $\bar{v}\neq 0$ is shown in blue. An unphysical fixed point with $\bar{v}\neq 0$ and $\bar{Y}=\bar{B}=\bar{\mu}=0$ is also present as a green dot at the bottom, but this is irrelevant for our purposes. The red plane at $\bar{v}=0$ on the left shows the accessible space of coupling parameters within the often used isotensional ensemble. Under fixed stress (isotensional) conditions, the thermal buckling transition occurs at $\sigma=0$ (i.e., $K=1$) and is controlled by the conventional vK$_{\rm{th}}$ fixed point. However, for fixed strain (isometric) conditions, when $\bar{v}\neq0$, we flow instead to a new codimension-1 fixed point (CvK$_{\rm th}$) that now controls the thermal buckling transition. The unstable renormalization group flow going towards large $K>1$ corresponds to strong compression and postbuckling behaviour. At low temperature, G$_\sigma$ is a globally attracting and stable fixed point which controls the properties of a tense flat membrane for both ensembles.}
	\label{fig:rgflow}
\end{figure*}

We now enumerate the four physically relevant fixed points \footnote{There are other interacting fixed points present, including two that correspond to a fluid membrane (with a vanishing shear modulus $\mu=0$) and a physically unrealizable one ($B=0$ but $v\neq 0$), that are not relevant for us.} permitted in both ensembles, to $\mathcal{\mathcal{O}}(\vare)$:
\begin{enumerate}[label=(\roman*)]
	\item G$_\sigma$: $K_*=0$, $\bar{Y}_*=0$, $\bar{\mu}_*=0$, $\bar{v}_*=0$.
	\item G$_\kappa$: $K_*=1$, $\bar{Y}_*=0$, $\bar{\mu}_*=0$, $\bar{v}_*=0$.
	\item vK$_{\rm{th}}$: $K_*=1$, $\bar{Y}_*=77\pi^2\vare/25$, $\bar{\mu}_*=96\pi^2\vare/25$, $\bar{v}=0$ ($\nu_p=-1/3$).
	\item CvK$_{\rm{th}}$: $K_*=1+\vare/50$, $\bar{Y}_*=77\pi^2\vare/25$, $\bar{\mu}_*=96\pi^2\vare/25$, $\bar{v}_*=16\pi^2\vare/25$ ($\nu_p=-1/3$).
\end{enumerate}
We have set $d_c=1$ here as is physically relevant; the expressions for the fixed points with arbitrary $d_c$ are presented in Appendix~\ref{app:rgd}.
Of these fixed points, only G$_\kappa$, G$_\sigma$ and vK$_{\rm{th}}$ are admissable in the isotensional ensemble. The thermal F{\"o}ppl-von K{\'a}rm{\'a}n fixed point vK$_{\rm{th}}$ has been the focus of virtually all studies to date. The \emph{constrained} thermal fixed point CvK$_{\rm{th}}$ is new and unique to the isometric ensemble.
Both G$_{\kappa}$ and G$_{\sigma}$ are Gaussian (noninteracting) fixed points that are bending and tension dominated respectively. The conventional flat phase is described by vK$_{\rm{th}}$, and occurs for a vanishing renormalized tension (hence $K_*=1$) that is appropriate for an unconstrained fluctuating membrane. This fixed point has been extensively studied previously \cite{nelson1987fluctuations,le1992self,aronovitz1988fluctuations,guitter1988crumpling,guitter1989thermodynamical,roldan2011suppression,kovsmrlj2016response} and it controls the buckling transition in the absence of boundary constraints, i.e., the \emph{isotensional} ensemble as evidenced by $\bar{v}=0$ at the fixed point.

In contrast, a new constrained fixed point CvK$_{\rm{th}}$ emerges in the isometric ensemble with $\bar{v}\neq 0$, reflecting the geometric constraint imposed by the clamped boundaries in the isometric ensemble. The new interacting fixed point involves bare compression (as $K_*>1$), unlike the others, reflecting the presence of a fluctuation induced spontaneous tension that appears only when the boundary is constrained. We will discuss this feature in more detail in Sec.~\ref{sec:buckling}. As we will show below, CvK$_{\rm{th}}$ controls the buckling transition in the \emph{isometric} ensemble.

\begin{figure*}[]
	\centering{\includegraphics[width=\textwidth]{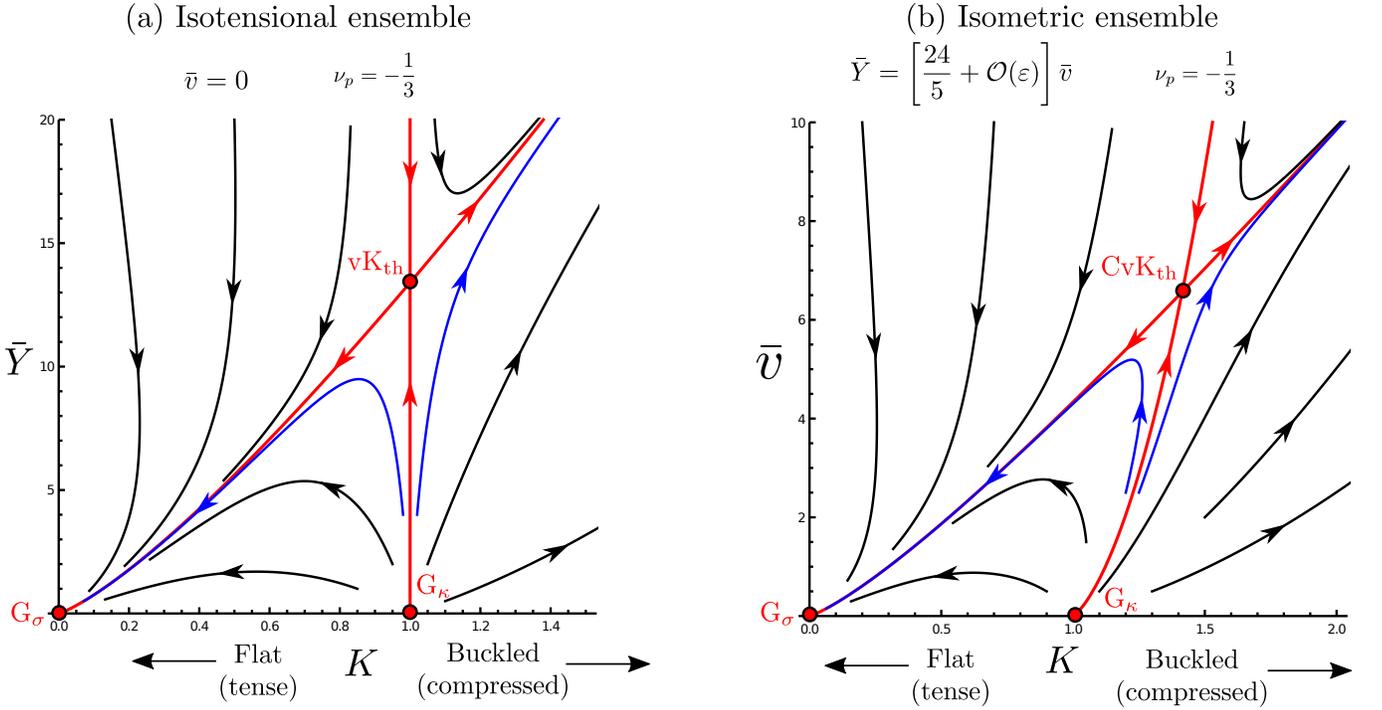}}
	\caption{A schematic of the renormalization group flows projected onto the invariant attracting planes appropriate to the two ensembles. The Poisson's ratio is fixed to its universal value $\nu_p=-1/3$ in both cases. (a) In the isotensional ensemble, $\bar{v}=0$ identically, and we have three fixed points, G$_\sigma$, G$_\kappa$ and vK$_{\rm{th}}$. The red lines are separatrices that connect the various fixed points and demarcate their basins of attraction. The vertical separatrix flowing into vK$_{\rm th}$ at $K=1$ decribes the buckling transition in this ensemble. The black streamlines are integral curves of the flow and the blue curves highlight two representative trajectories bracketing the buckling transition. (b) In the isometric ensemble, the relevant attractor is now a plane characterized by $\bar{Y}=z_*\bar{v}$ with $z_*=24/5+\mathcal{O}(\vare)$. Once again, we have three important fixed points, G$_\sigma$, G$_\kappa$ and CvK$_{\rm th}$. The red curves are separatrices delimiting stability basins for each fixed point and the separatrix flowing into CvK$_{\rm th}$ controls the buckling threshold in this ensemble. Unlike in the isotensional case, this line is curved and bends towards $K>1$, signalling the generation of spontaneous tension (Fig.~\ref{fig:buckling}). The black curves are streamlines of the local renormalization group flow and the blue curves are representative flow trajectories on either side of the buckling transition. In both ensembles, G$_{\sigma}$ is globally stable for flat and tense mebranes ($K<1$), while flows towards larger values of $K(>1)$ lead to strong compression and buckling.}
	\label{fig:flow}
\end{figure*}

We note that both vK$_{\rm{th}}$ and CvK$_{\rm{th}}$ are characterized by the universal Poisson's ratio $\nu_p=-1/3$. A schematic of the full renormalization group flow diagram with the above fixed points is sketched in Fig.~\ref{fig:rgflow}. The stability of each fixed point can be analyzed by linearizing about it. The two Gaussian fixed points, G$_\sigma$ and G$_\kappa$, differ simply by the presence or absence of $\sigma$ and play a role in both ensembles. The tension-dominated fixed point G$_\sigma$ is a globally stable attractor that controls the low temperature phase of a tense flat membrane. On the other hand, the bending dominated Gaussian fixed point G$_\kappa$ is unstable in all directions, as found previously for tense membranes \cite{guitter1989thermodynamical,roldan2011suppression,kovsmrlj2016response}. Both these fixed points along with their flow directions in different invariant planar sections of the full parameter space, are shown in Fig.~\ref{fig:flow}.

Within the invariant subspace of $\bar{v}=0$, associated with the isotensional ensemble (see Fig.~\ref{fig:rgflow}), the conventional flat phase fixed point vK$_{\rm{th}}$ has only one relevant direction corresponding to the external stress. As a result, it controls the finite temperature buckling transition in the isotensional ensemble, where the stress is tuned to zero at threshold. But once we allow for a $\bar{v}\neq 0$, i.e., work in the isometric ensemble instead, we find that vK$_{\rm{th}}$ is now a codimension two fixed point, being \emph{unstable} to this new nonlinear coupling. As sketched in Fig.~\ref{fig:rgflow}, the system can now flow to the new constrained fixed point CvK$_{\rm{th}}$ instead, which has codimension one. Perturbations in $\bar{Y},\bar{B}$ and $\bar{v}$ are all irrelevant at CvK$_{\rm{th}}$ and the only unstable or relevant direction is primarily along $K$. In other words, within the isometric ensemble, the strain tuned buckling transition is controlled by CvK$_{\rm{th}}$ and \emph{not} vK$_{\rm{th}}$.

The identification of two distinct ensemble-dependent fixed points controlling buckling is a significant achievement. The fact that the choice of fixed point is picked by the mechanical ensemble, here decided by fixed strain or stress boundary conditions, is quite intriguing. Although the isotensional and isometric ensembles are dual to each other, they remain inequivalent even in the thermodynamic limit, due to flexural phonons on all length scales at the critical point. As mentioned in the introduction, this remarkable feature is akin to Fisher renormalization of conventional critical exponents, which we demonstrate explicitly in Sec.~\ref{sec:scaling}. Below we compute the flat to buckled phase boundary and analyze the linearized flow in the vicinity of the two interacting fixed points to extract critical exponents for the buckling transition.

\section{Buckling transition in $4-\vare$ dimensions}
\label{sec:buckling}

Let us now analyze the recursion relations given in Eqs.~\ref{eq:Krg}-\ref{eq:nurg} in more detail.
%First we note that neither the bulk modulus $\bar{B}$ nor the Poisson's ratio $\nu_p$ enter the recursion relations for $K$, $\bar{Y}$ and $\bar{v}$.
For a given $\bar{Y}$, as $\bar{\mu}$ and $\nu_p$ are related, we only have to consider one of them. From Eq.~\ref{eq:nurg}, we easily see that the fixed point at $\nu_p=-1/3$ is stable and exponentially attracting for any finite $\bar{Y}$. So we shall neglect perturbations in the Poisson's ratio and fix $\nu_p=-1/3$. This condition in turn fixes the shear modulus to be $\bar{\mu}=(D+1)\bar{Y}/4$ and the bulk modulus to be $\bar{B}=(D+1)\bar{Y}/D(D+2)$, allowing us to then concentrate on the flow in the three-dimensional subspace of just $\{K,\bar{Y},\bar{v}\}$ parametrizing the stable attractor.

As an aside, note that $\nu_p$ rapidly approaches its fixed point value of $-1/3$ only when $\bar{Y}>0$, which is true in the vicinity of both the vK$_{\rm{th}}$ and CvK$_{\rm{th}}$ fixed points. In contrast, for a tense membrane governed by G$_{\sigma}$, $\bar{Y}(s)\propto e^{-s(4-\vare)}\to0$ approaches zero exponentially fast on large scales. In this case, $\nu_p$ does not reach its universal fixed point value and instead, the rapidly vanishing $\bar{Y}$ essentially freezes $\nu_p$ at a value that depends on microscopic properties of the material. Thus, while the large-distance Poisson's ratio is universal at the buckling transition, away from it, in a tense mebrane, it becomes nonuniversal and depends on microscopic details, consistent with results for fluctuating membranes under strong tension \cite{roldan2011suppression,kovsmrlj2016response,los2009scaling,burmistrov2018stress}. 

We shall now address buckling criticality in the two ensembles separately below.
\subsection{Isotensional ensemble}
In the isotensional ensemble, we set both $\sigma=\sigma_0$ and $\bar{v}=0$. The latter picks out a renormalization group invariant plane (see Figs.~\ref{fig:rgflow} and~\ref{fig:flow}a) specific to this ensemble. The buckling transition then occurs at $\sigma_0=\sigma_c=0$ (in the thermodynamic limit of infinite system size), though $\sigma_c$ is nonzero for a finite sheet (see Sec.~\ref{sec:mft}). Note that, right at the unconstrained fixed point vK$_{\rm{th}}$, we do have $\sigma_0=0$ (i.e., $K=1$) even at finite temperature, a result that holds to all orders in perturbation theory. Hence the critical stress at the buckling transition is still given by its $T=0$ value,
\begin{equation}
	\sigma_c(T)=-\kappa q_0^2\;,\label{eq:sigc}
\end{equation}
and it does \emph{not} receive corrections from thermal fluctuations in the isotensional ensemble. As before $q_0\sim 1/R$ is the smallest wavevector determined by the boundary conditions and the size of the sheet (Sec.~\ref{sec:mft}).

To compute anomalous scaling exponents at the transition, we use a standard renormalization group matching procedure \cite{rudnick1976equations} to relate correlation functions evaluated near the transition to those further away from the critical point. Under scaling, $\b{r}=b\b{r}'$ and $h(\b{r})=b^{\zeta}h(\b{r}')$, conversely, $\b{q}=\b{q}'/b$ and $h_{\b{q}}=h'_{\b{q}'}b^{D+\zeta}$ (in $D$-dimensions). Upon setting $b=e^s$, we have the following scaling relation for the height-height correlation function ($G_h(\b{q})\equiv\langle|h_{\b{q}}|^2\rangle/V_D$, where $V_D$ is the $D$-dimensional volume of the manifold)
\begin{equation}
	G_h(\b{q})=\exp\left\{\int_0^{s}\dd s'[D+2\zeta(s')]\right\}G_h(\b{q}e^{s};s)\;,
\end{equation}
where $G_h(\b{k};s)$ is computed using all the parameters evaluated at scale $s$. We now choose $s=s^*$ such that $|\b{q}|e^{s^*}=\ell_{\rm{th}}^{-1}$, set by the thermal length (Eq.~\ref{eq:lth}). This condition allows us to write,
\begin{align}
	G_h(\b{q})&=\exp\left[D s^*+2\int_0^{s^*}\dd s\;\zeta(s)\right]G_h(\ell_{\rm{th}}^{-1};s^*)\nonumber\\
	&=k_BT\ell_{\rm{th}}^{4}\dfrac{K(s^*)}{\kappa(s^*)}\exp\left[Ds^*+2\int_0^{s^*}\dd s\;\zeta(s)\right]\;.\label{eq:Cq}
\end{align}
Here we have used the fact that on small scales ($\ell\lesssim\ell_{\rm{th}}$), fluctuation corrections are negligible and a Gaussian or mean field treatment is valid. To evaluate the renormalized bending rigidity at scale $s^*$, we need the following flow equation as well
\begin{equation}
	\dfrac{\dd\kappa}{\dd s}=\kappa\left[2\zeta-\vare+\dfrac{5(\bar{Y}+4\bar{\mu})}{192\pi^2 K}\right]\;.\label{eq:kapparg}
\end{equation}
It is convenient to chose the height rescaling factor $\zeta(s)$ to keep $\kappa(s)$ fixed, which leads to
\begin{equation}
	\zeta(s)=\dfrac{\vare}{2}-\dfrac{5}{384\pi^2K(s)}\left[\bar{Y}(s)+4\bar{\mu}(s)\right]\;.
\end{equation}

%\begin{figure*}[]
%	\centering{\includegraphics[width=\textwidth]{fit.eps}}
%	\caption{Fitting the numericaly computed critical separatrix (red dots) to a functional form (black curve). (a) For $\bar{B}\to 0$, we find the logarithmic correction %to $K_c$ (Eq.~\ref{eq:Kcrit}). (b) For larger values of $\bar{B}$ and $K$, we fit a simple polynomial form to $\bar{B}_c(K)$ (Eq.~\ref{eq:Bcrit}).}
%	\label{fig:fit}
%\end{figure*}
Right at the buckling transition, the coupling constants flow to the fixed point vK$_{\rm{th}}$.
%We set $K$ to its fixed point value $K_*=1$ and let $\bar{Y}=\bar{Y}_*+\delta\bar{Y}$ ($\bar{Y}_*=77\pi^2\vare/25$), where small $\delta\bar{Y}$ perturbations relax rapidly back to the fixed point value as
%\begin{equation}
%	\dfrac{\dd\delta\bar{Y}}{\dd s}=-2\delta\bar{Y}\;.
%\end{equation}
%Thus $\bar{Y}(s)=\bar{Y}_*+\mathcal{O}(e^{-2s})$ in the vicinity of the vK$_{\rm{th}}$ fixed points.
The height correlator defines a renormalized bending rigidity via
\begin{equation}
	\kappa_R(\b{q})^{-1}=q^4\dfrac{G_h(\b{q})}{k_BT}\;.
\end{equation}
Upon using Eq.~\ref{eq:kapparg} and Eq.~\ref{eq:Cq} right at buckling, this then gives the well-known diverging bending rigidity,
\begin{equation}
	\kappa_R(\b{q})=\kappa(q\ell_{\rm{th}})^{-\eta}\;,\quad\eta=\dfrac{12}{25}\vare+\mathcal{O}(\vare^2)\;.\label{eq:eta1}
\end{equation}
The anomalous exponent $\eta$ that we obtain matches earlier calculations performed for unconstrained membranes \cite{aronovitz1988fluctuations,guitter1988crumpling}.
While we don't expect the one-loop approximation of $\eta$ to be numerically accurate in the physically relevant case of $D=2$ dimensions, we nonetheless obtain a reasonable value of $\eta=24/25\approx 0.96$ upon setting $\vare=2$ in Eq.~\ref{eq:eta1}. More sophisticated calculations involving self consistent \cite{le1992self,*le2018anomalous} and nonperturbative techniques \cite{kownacki2009crumpling} give $\eta\simeq 0.82-0.85$ which compares well with the exponent value measured in numerical simulations \cite{bowick1996flat,los2009scaling,roldan2011suppression,bowick2017non} and recent experiments \cite{Blees2015}.
%Although our one loop approximation is uncontrolled, the value of $\eta$ that we obtain is compatible with previous estimates and recovers the result of a previous fixed dimension calculation \cite{kovsmrlj2016response}.

The elastic moduli also experience a scale dependent renormalization, though they now get \emph{softer} on larger scales. The renormalized Young's modulus scales as
\begin{equation}
	Y_R(\b{q})=Y(q\ell_{\rm{th}})^{\eta_u}\;,\quad\eta_u=\dfrac{\vare}{25}+\mathcal{O}(\vare^2)\;,
\end{equation}
The other elastic moduli ($\mu,\lambda$ and $B$) all scale in the same way, with the same $\eta_u$ exponent. Both $\eta$ and $\eta_u$ are related via a Ward identity $\eta_u+2\eta=\vare$, which is a consequence of rotational invariance \cite{guitter1988crumpling,guitter1989thermodynamical,aronovitz1988fluctuations,aronovitz1989fluctuations,le1992self}.

If the external tension is small but nonzero, then we perturb slightly away from vK$_{\rm{th}}$. We write $K=1+\delta K$ and linearize in $\delta K$ to get
\begin{equation}
	\dfrac{\dd\delta K}{\dd s}=\left(2-\dfrac{12}{25}\vare\right)\delta K\equiv(2-\eta)\delta K\;,
\end{equation}
which is exactly true by virtue of the definition of $\eta$. For a small external stress ($|\sigma_0|\ll\kappa\Lambda^2$), $K\approx 1$, with $\delta K(s)=\delta K(0)e^{(2-\eta)s}$ growing with $s$, as expected of a relevant perturbation. Eventually, we reach a scale $s^*$ at which the external stress has grown large and is now comparable to the bending rigidity, i.e., $|\sigma(s^*)|\approx\kappa(s^*)\Lambda^2$), which defines the correlation length $\xi$ via $s^*=\ln(\xi/a)$ ($a\sim 1/\Lambda$ is a lattice cutoff) beyond which the stress dominates bending. After incorporating a nonzero $\sigma_c$ appropriate to a finite system, we have $\delta K(0)\propto|\sigma_0-\sigma_c|$ which gives
\begin{equation}
	\xi\propto|\sigma_0-\sigma_c|^{-\nu}\;,\quad\nu=\dfrac{1}{2-\eta}=\dfrac{1}{2}+\dfrac{3\vare}{25}+\mathcal{O}(\vare^2)\;,\label{eq:nu1}
\end{equation}
for the isotensional ensemble.
On short length scales ($\ell\ll\xi$), the system is controlled by the bending-dominated vicinty of the vK$_{\rm{th}}$ fixed point, while on larger scales ($\ell\gg\xi$) when $\sigma_0>0$, the system is dominated by external tension and the G$_{\sigma}$ fixed point. Hence, for $\sigma_0>0$, we have the following length scale dependence (for $D=2$)
\begin{equation}
	\dfrac{\kappa_R(\ell)}{\kappa}\propto\begin{cases}
		1\;,\quad(\ell<\ell_{\rm{th}})\\
		(\ell/\ell_{\rm{th}})^{\eta}\;,\quad(\ell_{\rm{th}}<\ell<\xi)\\
		(\xi/\ell_{\rm{th}})^{\eta}\ln(\ell/\xi)\;,\quad(\ell>\xi)
	\end{cases}\;,
\end{equation}
with similar expressions for the elastic moduli \cite{kovsmrlj2016response}. A similar dependence without the $\ln(\ell/\xi)$ term also exists for $2<D<4$ dimensions.

\subsection{Isometric ensemble}
Now that we have recapitulated the isotensional results, let us move onto the more interesting isometric ensemble. We now set $\sigma=B\e$ and identify $\bar{v}=\bar{B}$. As mentioned earlier, the $\bar{v}=0$ plane defining the isotensional ensemble is unstable to finite $\bar{v}$ perturbations, leading us to consider the full 3D space of parameters $\{K,\bar{Y},\bar{v}\}$. For $\bar{v}>0$, i.e., in the isometric ensemble now, we can further reduce dimensionality by writing $\bar{Y}=z\bar{v}$, which gives
\begin{equation}
	\dfrac{\dd z}{\dd s}=\dfrac{\bar{Y}}{16\pi^2}\left(1-\dfrac{5z}{24}\right)\;,\label{eq:z}
\end{equation}
to leading order in $\mathcal{O}(\vare)$.
With $\bar{Y}>0$ as before, this equation has an exponentially stable fixed point given by $z=24/5+\mathcal{O}(\vare)$ that attracts all the renormalization group flows for $\bar{v}>0$. This stable fixed point in $z$ hence defines an attracting invariant plane in the $\{K,\bar{Y},\bar{v}\}$ space (see Fig.~\ref{fig:flow}b), only accessible within the isometric ensemble. Note that the new constrained fixed point CvK$_{\rm{th}}$ lies on this plane as well, a welcome feature that guarantees that when $\bar{v}\neq 0$ microscopically and equal to $\bar{B}$ on short scales (as it should be in the isometric ensemble), their equality is retained on larger scales. More importantly, the relation $\bar{v}=\bar{B}>0$ serves as an invariant attractor under coarse-graining in the isometric ensemble. Within this plane, we have two coupled flow equations, namely
\begin{align}
	\dfrac{\dd K}{\dd s}&=2K(K-1)+\dfrac{\bar{v}}{16\pi^2}(12-13K)\;,\\
	\dfrac{\dd\bar{v}}{\dd s}&=\bar{v}\left[\vare+4(K-1)-\dfrac{27\bar{v}}{16\pi^2}\right]\;,
\end{align}
as shown in Fig.~\ref{fig:flow}b.

Upon dividing the above two equations, we obtain $\dd K/\dd\bar{v}$ which we numerically integrate to obtain the basin of attraction of G$_\sigma$ and CvK$_{\rm{th}}$. The stable and unstable manifolds are obtained as integral curves of the stable and unstable eigendirections at CvK$_{\rm{th}}$, which correspond to separatrices shown in red in Fig.~\ref{fig:flow}b. The separatrix connects the unstable fixed point G$_\kappa$ at $K=1$ to the constrained thermal buckling transition fixed point at CvK$_{\rm{th}}$, and it delineates the stability region for a clamped membrane. All parameter values that fall to the left of this separatrix flow into the stable G$_\sigma$ fixed point, leading to a sheet that is flat and tense on large scales. In the opposite case, parameter values starting to the right of the red separatrix flow away to larger values of $K$, signalling strong compression ($\sigma=B\e<0$) and buckling of the membrane. Representative flow trajectories illustrating this are shown in blue in Fig.~\ref{fig:flow}b. The separatrices are computed as the solution of a boundary value problem and hence don't admit a simple analytical solution. Nonetheless, we can obtain some asymptotic results informed by the algebraic structure of the recursion relations, in conjunction with the numerical solution.

\begin{figure}[]
	\centering{\includegraphics[width=0.5\textwidth]{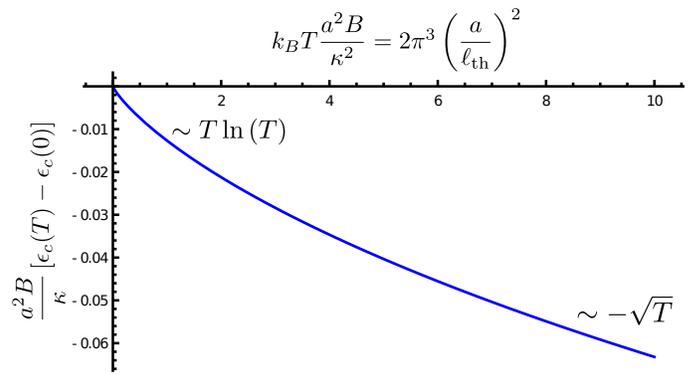}}
	\caption{The numerically computed buckling threshold $|\e_c(T)|$ extrapolated to $\vare=2$ ($D=2$). The critical buckling strain gets shifted to more negative (compressive) values at finite temperature due to the generation of a thermally induced spontaneous tension. We have subtracted out the zero temperature buckling threshold $\e_{c}(0)=-\kappa q_0^2/B$. Near the buckling transition, we have used the universal Poisson's ratio ($\nu_p=-1/3$) to relate the bulk and Young's modulus in $D=2$. At low temperature (large $\ell_{\rm th}$), $\e_c(T)\sim T\ln T$, while at higher temperature (smaller $\ell_{\rm th}$), we have $\e_c(T)\sim-\sqrt{T}$ (Eq.~\ref{eq:ecT}). While this plot is obtained by setting $\vare=2$ in the resursion relations obtained by expanding around $D=4$ dimensions, a qualitatively similar curve is obtained from a fixed dimension calculation with $D=2$ and $d=3$ (Appendix~\ref{app:rg}).}
	\label{fig:buckling}
\end{figure}
Upon using the definitions of $K$ and $\bar{B}$ and the relation $\sigma=B\e$, we obtain the critical curve for the buckling transition strain ($\e_c(T)$) in terms of the elastic constants and temperature, as plotted in Fig.~\ref{fig:buckling} for $D=2$ ($\vare=2$). For low and high temperatures, we find simple asymptotic expansions for the buckling threshold $\e_c(T)<0$ (in $D=2$),
\begin{subequations}
\begin{align}
	|\e_c(T)|&\simeq\dfrac{\kappa q_0^2}{B}+\dfrac{k_BT}{8\pi\kappa}\left[2\ln\left(\dfrac{a}{\ell_{\rm{th}}}\right)+c_1\right]\;,\quad(\ell_{\rm{th}}\gg a)\;,\\
	|\e_c(T)|&\simeq\dfrac{\kappa q_0^2}{B}+\dfrac{k_BT}{8\pi\kappa}\left(\dfrac{\ell_{\rm{th}}}{a}\right)\left[c_2-c_3\dfrac{\ell_{\rm{th}}}{a}\right]\;,\quad(\ell_{\rm{th}}\ll a)\;,
\end{align}\label{eq:ecT}
\end{subequations}
where $a\sim\Lambda^{-1}$ is the lattice cutoff and $c_{1,2,3}$ are numerical constants that must be computed by numerical integration of the recursion relations. While here, we extrapolated our perturbative solution to $\vare=2$, we have confirmed that the same asymptotic expressions for the buckling strain, with only $c_{1,2,3}$ modified, are also obtained within a fixed dimension calculation with $D=2$ and $d=3$ (Appendix~\ref{app:rg}). As $T\to 0$, $\e_c(T)\to-\kappa q_0^2/B$, which is the zero temperature buckling instability threshold with $q_0\sim 1/R$ being the smallest available mode in a system of size $R$ (Sec.~\ref{sec:mft}). We have utilized the fact that, near the transition $\nu_p=-1/3$, which relates $Y$ and $B$ via $B=3Y/8$ (in $D=2$), allowing us to write the above in terms of the thermal length $\ell_{\rm{th}}$.
As $\ell_{\rm{th}}\sim T^{-1/2}$ (Eq.~\ref{eq:lth}), for $D=2$ we have $|\e_c(T)|\sim T\ln(1/T)$ at low temperature and $|\e_c(T)|\sim\sqrt{T}$ at high temperature, as shown in Fig.~\ref{fig:buckling}.
For general $D$, as $T\to 0$, we find that $|\e_c(T)|\sim T[1+{\rm const.}\;(T^{2/\vare-1}-1)/(2-\vare)]$ and the linear $T$ dependence dominates at small $T$ for $0<\vare<2$ ($2<D<4$), while an additional logarithmic term $\sim T\ln(T)$ appears when $\vare=2$ ($D=2$). Note that the high temperature asymptotics depends only weakly on dimension.

We pause here to comment on the above results. Unlike in the isotensional ensemble, where the buckling threshold did \emph{not} receive any correction from thermal fluctuations (Eq.~\ref{eq:sigc}), in the isometric ensemble, the buckling threshold gets pushed to higher values of compression (as $\e_c(T)<0$ and $|\e_c(T)|$ increases with $T$) at higher temperature. In other words, the sheet spontaneously develops a tension due to thermal fluctuations in the isometric ensemble. A freely fluctuating sheet wants to shrink in-plane for entropic reasons, but the clamped boundaries resist this shrinkage, thereby putting the sheet under tension. As a result, the externally imposed strain now has to compensate and overcome this thermally induced tension in order to cause buckling. This effect is absent in the isotensional ensemble because the boundaries are free to fluctuate, allowing the sheet to freely shrink with increasing temperature, albeit against a constant external stress.

We now compute critical scaling exponents at the buckling transition. Here, by tuning right to the buckling threshold, we approach the constrained fixed point CvK$_{\rm{th}}$. Upon evaluating the height-height correlator, we obtain the renormalized bending rigidity to be
\begin{equation}
	\kappa_R(\b{q})=\kappa(q\ell_{\rm{th}})^{-\eta'}\;,\quad\eta'=\dfrac{12}{25}\vare+\mathcal{O}(\vare^2)\;.
\end{equation}
Remarkably, we obtain the same anomalous scaling exponent here as in the isotensional ensemble (Eq.~\ref{eq:eta1}). As we will show later in Sec.~\ref{sec:exp}, we expect $\eta=\eta'$ from general scaling arguments, which we also confirm for arbitrary $d_c$ within a lowest order systematic $\vare=4-D$ expansion in Appendix~\ref{app:rgd} and Table~\ref{tab:exponents}.

A similar analysis of the phonon correlator or equivalently the nonlinear stretching term also provides the renormalized Young's modulus,
\begin{equation}
	Y_R(\b{q})=Y(q\ell_{\rm{th}})^{\eta'_u}\;,\quad\eta'_u=\dfrac{\vare}{25}+\mathcal{O}(\vare^2)\;.
\end{equation}
As before, $\eta'_u$ satisfies the Ward identity $\eta'_u+2\eta'=\vare$. A consequence of the equality $\eta=\eta'$ is that $\eta_u=\eta'_u$ as well, which is verified here to leading order in an expansion in $\vare=4-D$.

Distinct critical exponents appear, however, when we perturb away from the buckling threshold, with one relevant direction that flows away from the fixed point CvK$_{\rm{th}}$. Upon linearizing about this fixed point, we obtain and diagonalize the resulting Jacobian matrix to produce the following eigenvalues valid to $\mathcal{O}(\vare)$,
\begin{equation}
	y_0=2-\dfrac{13\vare}{25}\;,\ y_1=y_2=-\dfrac{\vare}{25}\;,\ y_3=-\vare\;.
\end{equation}
We have three irrelevant directions with negative eigenvalues ($y_{1,2,3}<0$) and one relevant direction with a positive eigenvalue ($y_0>0$). If we write $K(s)=K_*+\delta K(s)$, where $K_*=1+\vare/50$ is the fixed point value and $\delta K(s)$ is a small deviation, then we find that $\delta K(s)\simeq \delta K(0)e^{y_0s}$ grows with scale as the renormalization group flow proceeds away from CvK$_{\rm{th}}$ along the outgoing separatrix.
Note that $\delta K(0)\propto (\e_c(T)-\e)$ is controlled by the distance to the buckling transition. This relation can be easily obtained by expanding $\sigma/(\kappa\Lambda^2)=K^{-1}-1$ to linear order in $\delta K$ and setting $\sigma=B\e$ as appropriate in the isometric ensemble. Upon using standard renormalization group arguments, we obtain the divergent correlation length to be
\begin{equation}
	\xi\propto|\e-\e_c(T)|^{-\nu'}\;,\quad\nu'=\dfrac{1}{y_0}=\dfrac{1}{2}+\dfrac{13}{100}\vare+\mathcal{O}(\vare^2)\;.
\end{equation}
Strikingly, we obtain a distinct value of $\nu'$ here as compared to the value of $\nu$ obtained in the isotensional ensemble. In Sec.~\ref{sec:exp}, we in fact demonstrate through general scaling arguments that $\nu<2/D<\nu'$, which is satisfied to leading order in $\vare$ in our systematic expansion (see also Table~\ref{tab:exponents}). While, we don't expect this leading order perturbative calculation to remain numerically accurate for $\vare=2$, the exponent inequality continues to hold in $D=2$, where it reduces to $\nu<1<\nu'$. Note that, the requirement $\nu'>1$ for $D=2$ in the isometric ensemble implies an exceptionally strong divergence for a correlation length in critical phenomena.
%Eventually we reach $s=s^*$ at which point $\delta K$ has increased to a finite value, say $\delta K(s^*)=0.1$, beyond which, critical fluctuation effects are suppressed and a simple Gaussian description suffices as the flow is eventually captured by G$_{\sigma}$. The precise choice of this value does not affect the critical exponents. The correlation length is then given by $\xi=a\;e^{s^*}$. By using $\delta K(s)\propto\Delta(s)=\Delta_0\;e^{y_1s}$ along with $|\Delta_0|\propto|\e-\e_c(T)|$, we obtain

The difference in exponents ($\nu\neq\nu'$) directly demonstrates that the universality class for buckling within the isometric ensemble is distinct from that in the isotensional ensemble, as advertised in the introduction.

Our analysis of the renormalization group flow at the two fixed points associated with buckling in the two dual ensembles is now complete. While the calculation presented here was performed within a systematic $\vare=4-D$ expansion at fixed manifold codimension $d_c$, we provide the general results for arbitrary $d_c$ in Appendix~\ref{app:rgd}. We also present a simpler, yet uncontrolled, and hence inaccurate one-loop approximation for the fixed points and exponents directly in fixed internal and embedding dimension ($D=2,d=3$) in Appendix~\ref{app:rg}.
Although, we directly compute only the scaling exponents associated with the fluctuation spectra ($\eta,\eta'$ or equivalently $\eta_u,\eta'_u$) and the correlation length ($\nu,\nu'$), the other exponents defined in Sec.~\ref{sec:exp} can be obtained through various exponent identities derived below.
All the exponents for both ensembles are listed to leading order in an $\vare=4-D$ expansion for arbitrary codimension $d_c=d-D$ in Table~\ref{tab:exponents}. We also use the most accurate estimates for the $\eta$ exponent in the physically relevant dimensions of $D=2$ and $d=3$, obtained through the self-consistent screening approximation \cite{le1992self} along with the scaling relations derived in Sec.~\ref{sec:scaling} to directly quote our best estimates for the various buckling exponents in both ensembles, in Table~\ref{tab:exponents}.

Below, we present a general scaling theory valid near the buckling transition and derive relations between various exponents, which acquire nonstandard forms due to the unusual size dependence exhibited by the buckling transition. This framework will also allow us to explicitly demonstrate that the distinction between the isotensional and isometric buckling universality classes constitutes a mechanical variant of Fisher renormalization \cite{fisher1968renormalization}.

\section{Scaling Relations \& Fisher Renormalization}
\label{sec:scaling}
In this section, we continue to work in the general setting of a $D$-dimensional elastic manifold embedded in $d$-dimensional Euclidean space. As before, the codimension $d_c=d-D>0$ counts the number of directions in which the elastic material can deform extrinsically, i.e., the flexural modes. Close to the thermalized buckling transition, we have universal scaling laws, just as in conventional critical phenomena, compactly captured by the scaling form of the free energy itself. Standard renormalization group arguments show that the free energy density defined by $F=-(k_BT/V_D)\ln[\int\mathcal{D}h\;e^{-\mathcal{F}/k_BT}]$ ($V_D$ is the $D$-dimensional volume, which is just $V_2=A$ the area for $D=2$) has a singular part $F_s$ which obeys the following scaling relation close to the transition \cite{goldenfeld2018lectures}
\begin{equation}
	F_s=b^{-D}\tilde{\Psi}_F\left(\Delta\sigma b^{1/\nu},\mathcal{E}b^{y_{\mathcal{E}}}\right)\;,\label{eq:Fs1}\\
\end{equation}
where $b$ is a scale factor and $\tilde{\Psi}_F$ is a scaling function that implicitly depends on the system size via $R/b$, the bending rigidity via $\kappa b^{-\eta}$ and the elastic moduli via $\{Y,B\}b^{\eta_u}$. We suppress this dependence to ease notation, but these quantities are important as they give rise to nonstandard scaling relations later. Eq.~\ref{eq:Fs1} allows us to map the physics near the buckling transition onto the mean field theory derived in Sec.~\ref{sec:mft}. A finite external field $\mathcal{E}$ is a strongly relevant perturbation, and has been retained with its scaling exponent $y_{\mathcal{E}}>0$. For convenience, we will work within the isotensional ensemble where the distance from the buckling threshold is given by $\Delta\sigma=\sigma_0-\sigma_c(T)$ \footnote{Note that $\sigma_c(T)$ is in general size dependent in a finite sheet, due to the zero temperature buckling threshold $\propto 1/R^2$, see Sec.~\ref{sec:mft}. In addition, fluctuation effects will lead to a shift in the buckling threshold $\propto 1/R^{1/\nu}$ as per standard finite size scaling arguments \cite{amit2005field}, which we expect to dominate over the zero temperature threshold in a large sheet. A similar argument holds in the isometric ensemble as well.}. Equivalent results for the isometric ensemble will be quoted directly as they follow immediately by replacing $\Delta\sigma$ with $B\Delta\e=B(\e-\e_c(T))$ and exchanging the unprimed exponents for the primed ones. This connection holds for all the exponent identities, except for the stress-strain exponents $\theta$ and $\theta'$, which require a minor modification due to their definition (Eq.~\ref{eq:theta}), as will be clear later on.

By choosing $b=|\Delta\sigma|^{-\nu}\propto\xi$, we scale out the $\Delta\sigma$ dependence to obtain
\begin{equation}
	F_s=|\Delta\sigma|^{\nu D}\Psi_F\left(\dfrac{\mathcal{E}}{|\Delta\sigma|^{\phi}}\right)\;,\label{eq:Fsing}
\end{equation}
where $\Psi_F(x)=\tilde{\Psi}_F(1,x)$. The crossover or gap exponent is given by $\phi=\nu y_{\mathcal{E}}$ (correspondingly $\phi'=\nu'y'_{\mathcal{E}}$ in the isometric ensemble). The full scaling form,
\begin{equation}
	F_s=|\Delta\sigma|^{\nu D}\Psi_F\left(\dfrac{\mathcal{E}}{|\Delta\sigma|^{\phi}},\kappa|\Delta\sigma|^{\nu\eta},\dfrac{\{Y,B\}}{|\Delta\sigma|^{\nu\eta_u}},R|\Delta\sigma|^{\nu}\right)\;,
\end{equation}
is a function of five distinct variables, four of which we have suppressed in Eq.~\ref{eq:Fsing}.

The height field has a scaling dimension $\zeta$. Right at buckling, we expect $\langle h(\b{r})^2\rangle\sim\int_{1/R}\dd^Dq /(\kappa_R(\b{q})q^4)\sim R^{2\zeta}$, which gives \cite{aronovitz1988fluctuations,guitter1988crumpling,guitter1989thermodynamical,aronovitz1989fluctuations,le1992self}
\begin{equation}
	\zeta=\dfrac{4-D-\eta}{2}\;,\quad\zeta'=\dfrac{4-D-\eta'}{2}\;.\label{eq:zeta}
\end{equation}
Similarly, the requirement that the rotationally invariant nonlinear strain tensor $u_{ij}$ renormalize correctly leads to a Ward identity exponent relation \cite{guitter1988crumpling,guitter1989thermodynamical,aronovitz1988fluctuations,aronovitz1989fluctuations,le1992self}, now stated in general $D$
\begin{equation}
	2\eta+\eta_u=4-D\;,\quad2\eta'+\eta'_u=4-D\;.\label{eq:ward}
\end{equation}
These are well known identities, which we will use below in deriving additional exponent relations.

Similar to the mean field treatment of buckled ribbons in Ref.~\cite{hanakata2020thermal}, the average height $\langle h\rangle$ serves as an order parameter for the buckling transition here. By definition, we obtain (for general $\mathcal{E}$)
\begin{equation}
	\langle h\rangle=-\dfrac{\partial F_s}{\partial\mathcal{E}}=|\Delta\sigma|^{\nu D-\phi}\Psi_h\left(\dfrac{\mathcal{E}}{|\Delta\sigma|^{\phi}}\right)\label{eq:h}
\end{equation}
where $\Psi_h(x)=\Psi'_F(x)$. Now, we identify $\langle h\rangle\sim b^{\zeta}\propto|\Delta\sigma|^{-\nu\zeta}$, which gives the exponent relations
\begin{equation}
	\phi=\dfrac{\nu}{2}(4+D-\eta)\;,\quad\phi'=\dfrac{\nu'}{2}(4+D-\eta')\;.\label{eq:phi}
\end{equation}
As a consistency check, one can easily confirm that this relation for the gap exponents is equivalent to demanding a trivial renormalization of the external field $\mathcal{E}$ (see Eq.~\ref{eq:dEdl} in Sec.~\ref{sec:rg} for the $D=2$ version).

In the zero field limit ($\mathcal{E}=0$), for strong compression, we spontaneously break symmetry by buckling, which leads to a finite $\langle h\rangle$. As $\Delta\sigma\to 0$ in a finite system, the rescaled bulk and Young's moduli ($B$ and $Y$) diverge as $|\Delta\sigma|^{-\eta_u\nu}$, while the rescaled system size ($R|\Delta\sigma|^{\nu}$) and bending rigidity ($\kappa|\Delta\sigma|^{\nu\eta}$) become vanishingly small. But the latter can't be set to zero quite yet. We know from our mean field analysis that upon buckling, $\langle h\rangle\propto R/Y^{1/2}$ (Eq.~\ref{eq:H0} in Sec.~\ref{sec:mft}). As a result, the elastic moduli act as relevant scaling variables at the transition while the system size behaves as a \emph{dangerously irrelevant variable} \cite{amit1982dangerous} whose scaling nontrivially affects the critical exponents. For $\mathcal{E}=0$, this observation leads to
\begin{equation}
	\lim_{\Delta\sigma\to 0}\Psi_h(0)\propto\dfrac{R}{Y^{1/2}}|\Delta\sigma|^{\nu(1+\eta_u/2)}\;.\label{eq:psi0}\\
\end{equation}
Note that this interesting size dependence for the scaling function is not a statement of conventional finite-size scaling \cite{amit2005field}; it is instead a unique feature arising from the long wavelength nature of the buckling transition.
By combining Eqs.~\ref{eq:zeta},~\ref{eq:ward} with this asymptotic behaviour, we can compute the order parameter exponents $\beta,\beta'$ for the buckled height (Eq.~\ref{eq:beta}) in our two ensembles to be
\begin{equation}
	\beta=\nu\left(1-\dfrac{\eta}{2}\right)\;,\quad\beta'=\nu'\left(1-\dfrac{\eta'}{2}\right)\;.\label{eq:beta1}
\end{equation}
This exponent identity is new and distinct from the usual hyperscaling relation that relates $\beta$ and $\eta$ in conventional critical phenomena (the latter in our current notation would read as $\beta=-\nu\zeta<0$, which is obviously wrong).

We can similarly compute the susceptibility exponents $\gamma,\gamma'$ defined in Sec.~\ref{sec:exp}. From our mean field analysis, we know that $\chi\propto R^2$ (Eq.~\ref{eq:chi}). Upon taking a derivative of Eq.~\ref{eq:h} and evaluating it at $\mathcal{E}=0$, we obtain
\begin{equation}
	\lim_{\Delta\sigma\to 0}\Psi'_h(0)\propto R^2|\Delta\sigma|^{2\nu}\;.\label{eq:psi10}\\
\end{equation}
The susceptibility scaling exponents then satisfy
\begin{equation}
	\gamma=\nu(2-\eta)\;,\quad\gamma'=\nu'(2-\eta')\;.\label{eq:gamma1}
\end{equation}
Although we recover the standard Fisher's identity, its appearance is in fact nontrivial, as is easily seen by noting that Fisher's identity reflects the equilibrium fluctuation-response relation \cite{chaikin1995principles},
\begin{equation}
	k_BT\chi=\int\dd^Dr\langle h(\b{r})h(\b{0})\rangle\;,
\end{equation}
which by a na{\"i}ve application of scaling would give $\chi\sim\xi^{4-\eta}$ resulting in $\gamma=\nu(4-\eta)$ instead of Eq.~\ref{eq:gamma1}. The resolution, as before, lies in the nontrivial size dependence of the correlation function integral (because $h(\b{r})\sim R$), which leads to $k_BT\chi\sim\xi^{4-\eta}(R/\xi)^2\propto\xi^{2-\eta}$ and then correctly producing Eq.~\ref{eq:gamma1}. In addition, combining Eqs.~\ref{eq:beta1} and~\ref{eq:gamma1} results in another unusual exponent identity
\begin{equation}
	\gamma=2\beta\;,\quad\gamma'=2\beta'\;.\label{eq:g2b}
\end{equation}
Note that the inequality $\nu\neq \nu'$ (discussed below) will necessarily imply additional exponent differences between the two ensembles, such as $\phi\neq\phi'$, $\beta\neq\beta'$ and $\gamma\neq\gamma'$ (see Table~\ref{tab:exponents} for a summary).

Next we move on to the nonlinear response in the presence of an external field. For finite $\mathcal{E}$ we now take the limit $\Delta\sigma\to 0$, which requires that we focus on the $x\to\infty$ asymptotics of $\Psi_h(x)\sim x^{1/\delta}$. Once again, we must be careful with regard to scaling with the elastic moduli and the system size. The mean field analysis (Eq.~\ref{eq:H0E}) dictates that the prefactor to the nonlinear response itself scales as $(R^4/Y)^{1/3}$. Upon taking this dependence into account, we have
\begin{equation}
	\lim_{\Delta\sigma\to 0}\Psi_h\left(\dfrac{\mathcal{E}}{|\Delta\sigma|^{\phi}}\right)\propto\left(\dfrac{\mathcal{E}}{|\Delta\sigma|^{\phi}}\right)^{1/\delta}\left(\dfrac{R^4|\Delta\sigma|^{4\nu}}{Y|\Delta\sigma|^{-\nu\eta_u}}\right)^{1/3}\;,\label{eq:psiinf}\\
\end{equation}
By demanding that the $\Delta\sigma$ dependence cancel as $\Delta\sigma\to 0$, we determine $\delta$ and $\delta'$. Remarkably, upon using the other identites from Eqs.~\ref{eq:zeta},~\ref{eq:ward} and~\ref{eq:phi}, we obtain
\begin{equation}
	\delta=\delta'=3\;,\label{eq:delta1}
\end{equation}
which is an exact result, independent of dimension! In Appendix~\ref{app:simple}, we show how the above results can all be combined into a simpler scaling form for $\langle h\rangle$ with a \emph{single} size dependent scaling variable and a modified gap exponent.

\begin{table*}[t]
	{\centering
	{\setlength{\extrarowheight}{4pt}
	\begin{tabular*}{\textwidth}{@{\extracolsep{\fill} }cccccc}
		\hline
		\hline
		\multirow{2}{*}{Exponent} & \multirow{2}{*}{Relation} &\multicolumn{2}{c}{Isotensional ensemble} & \multicolumn{2}{c}{Isometric ensemble}\\
		& & $D=2,\,d=3^{\dagger}$ & $D=4-\vare,\,d=D+d_c$ & $D=2,\,d=3^{\dagger}$ & $D=4-\vare,\,d=D+d_c$\\[2pt]
		\hline
		\hline
		\rule{0pt}{5ex}
		$\eta,\,\eta'$ & $\eta'=\eta$  & $0.821$ & $\dfrac{12\vare}{24+d_c}$ & $0.821$ & $\dfrac{12\vare}{24+d_c}$\\[1em]
		$\eta_u,\eta'_u$ & $\begin{aligned}2\eta+\eta_u&=4-D\\ 2\eta'+\eta'_u&=4-D\end{aligned}$ & $0.358$ & $\dfrac{d_c\vare}{24+d_c}$ & $0.358$ & $\dfrac{d_c\vare}{24+d_c}$\\[1.5em]
			$\nu,\nu'$ & $\begin{aligned}\nu&=1/(2-\eta)\\ \nu'&=\nu/(D\nu-1)\end{aligned}$ & $0.848$ & $\dfrac{1}{2}+\dfrac{3\vare}{24+d_c}$ & $1.218$ & $\dfrac{1}{2}+\dfrac{(12+d_c)\vare}{(96+4d_c)}$\\[1.5em]
		$\beta,\beta'$ & $\begin{aligned}\beta&=\nu(1-\eta/2)\\ \beta'&=\nu'(1-\eta'/2)\end{aligned}$ & $\dfrac{1}{2}$ & $\dfrac{1}{2}$ & $0.718$ & $\dfrac{1}{2}+\dfrac{d_c\vare}{96+4d_c}$\\[1.5em]
		$\gamma,\gamma'$ & $\begin{aligned}\gamma&=\nu(2-\eta)=2\beta\\ \gamma'&=\nu'(2-\eta')=2\beta'\end{aligned}$ & $1$ & $1$ & $1.436$ & $1+\dfrac{d_c\vare}{48+2d_c}$\\[1.5em]
		$\delta,\delta'$ & $\begin{aligned}\delta\beta&=\gamma+\beta\\ \delta'\beta'&=\gamma'+\beta'\end{aligned}$ & $3$ & $3$ & $3$ & $3$\\[1.5em]
		$\theta,\theta'$ & $\begin{aligned}&\theta=1/(\nu D-1)\\ \theta'&=\nu'D-1,\,\theta'=\theta\end{aligned}$ & $1.436$ & $1+\dfrac{d_c\vare}{48+2d_c}$ & $1.436$ & $1+\dfrac{d_c\vare}{48+2d_c}$\\[1.5em]
		$\phi,\phi'$ & $\begin{aligned}\phi&=\nu(4+D-\eta)/2\\ \phi'&=\nu'(4+D-\eta')/2\end{aligned}$ & $2.196$ & $2+\dfrac{(12-d_c)\vare}{96+4d_c}$ & $3.155$ & $2+\dfrac{3(4+d_c)\vare}{4(24+d_c)}$\\[1.5em]
		\hline
		\hline
	\end{tabular*}}
	}
	\raggedright
	\footnotesize{$^\dagger$ Exponents computed using $\eta=4/(1+\sqrt{15})$ obtained from the self-consistent screening approximation \cite{le1992self}.}
	\caption{The two buckling universality classes. We list all the scaling exponents along with the relevant scaling identites for the isotensional (unprimed exponents) and isometric (primed exponents) ensembles. The exponents are obtained within an $\vare$-expansion ($D=4-\vare,d=D+d_c$), accurate to $\mathcal{O}(\vare)$ and also in physical dimensions ($D=2,d=3$) using the best self-consistent estimates for $\eta$ \cite{le1992self}. Only $\beta=1/2$, $\gamma=1$ (isotensional) and $\delta=\delta'=3$ (both ensembles) are exact to all orders and independent of dimensionality.}
	\label{tab:exponents}
\end{table*}

Finally, we study the anomalous stress-strain curve exponents $\theta,\theta'$ defined by Eqs.~\ref{eq:theta}. These quantities are simpler as they approach finite limits when the system size $R\to\infty$. Here we distinguish the isotensional and the isometric ensemble as $\theta$ and $\theta'$ are defined differently in the two. We now note that the scaling variable is $B\Delta\e$ in the isometric ensemble, set $\mathcal{E}=0$ in Eq.~\ref{eq:Fsing} and employ the definitions in Eqs.~\ref{eq:es},~\ref{eq:se} to get
\begin{subequations}
\begin{align}
	\langle\e\rangle&=-\dfrac{\partial F_{s,\sigma}}{\partial\sigma_0}\propto|\Delta\sigma|^{\nu D-1}\sim|\Delta\sigma|^{1/\theta}\quad(\textrm{isotensional})\;,\\
	\dfrac{\langle\sigma\rangle}{B}&=\dfrac{\partial F_{s,\e}}{\partial(B\e)}\propto|B\Delta\e|^{\nu'D-1}\sim|\Delta\e|^{\theta'}\quad(\textrm{isometric})\;.
\end{align}
\end{subequations}
In this mechanical context $\theta,\theta'$ take on the role usually played by energy scaling in conventional critical phenomena. We now obtain exponent relations analogous to Josephson's hyperscaling relation \cite{goldenfeld2018lectures}
\begin{equation}
	\theta=\dfrac{1}{\nu D-1}\;,\quad\theta'=\nu'D-1\;.\label{eq:theta1}
\end{equation}
We shall see later that $\theta,\theta'>1$ (see Table~\ref{tab:exponents}), which leads to a crucial distinction between the two ensembles. In the isotensional ensemble, for $1/\theta<1$, the anomalous sublinear response dominates any linear Hookean response as $\Delta\sigma\to 0$ \cite{kovsmrlj2016response}. In contrast, in the isometric ensemble, for $\theta'>1$, the dominant strain response is in fact the nonsingular linear term as $\Delta\e\to 0$. This dichotomy reflects a crucial physical consequence of the different boundary conditions: in the isotensional ensemble, the sheet is infinitely compliant to homogeneous dilations or contractions in the plane, which is a zero mode of the system, but in the isometric ensemble, the clamped boundary conditions prohibit this zero mode and the sheet has a finite compliance to homogeneous isotropic distortions.

We now discuss one more scaling relation that is \emph{only} true in the isotensional ensemble and \emph{not} in the isometric ensemble. In the isotensional case, as we saw before in Sec.~\ref{sec:rg}, the in-plane tension $\sigma=\sigma_0$ does not receive any graphical corrections as $\bar{v}=0$ identically. This is true in any dimension and to all orders in perturbation theory, as a consequence of the fact that $\sigma_0$ is the sole term that breaks rotational invariance, while the bending and nonlinear stretching terms preserve rotational symmetry. This nonrenormalization condition then implies
\begin{equation}
	\nu=\dfrac{1}{2-\eta}\;,\label{eq:nu2}
\end{equation}
as we already saw by explicit calculation in Eq.~\ref{eq:nu1}. Because we must have $\eta,\eta_u>0$, Eq.~\ref{eq:ward} implies that $\eta\leq(4-D)/2$ (for $D\leq4$). This inequality then shows that $\nu\leq 2/D$ always.
An additional important consequence of Eq.~\ref{eq:nu2} is that both $\beta$ and $\gamma$ take on their mean-field values,
\begin{equation}
	\beta=\dfrac{1}{2}\;,\quad\gamma=1\;,
\end{equation}
in any dimension.
Note that such relations do \emph{not} hold in the isometric ensemble as can be seen in an $\vare$-expansion as displayed in Table~\ref{tab:exponents}. Eq.~\ref{eq:nu2} determining $\nu$ when plugged into the hyperscaling relation (Eq.~\ref{eq:theta1}) also gives $\theta=(2-\eta)/(D-2+\eta)$, which is consistent with the scaling expected from $\langle\e\rangle\sim\int(\dd^Dr/V_D)\langle|\vec{\del}h|^2\rangle\sim\xi^{2\zeta-2}$. This result agrees with previously derived scaling relations in arbitrary dimensions \cite{guitter1989thermodynamical,aronovitz1988fluctuations} and leads to $\theta=(2-\eta)/\eta$ when $D=2$ \cite{kovsmrlj2016response}.

This completes the derivation of the various scaling exponents in the two ensembles. The values of the exponents computed within an $\vare$-expansion ($D=4-\vare,d=D+d_c$)and by using estimates from a self-consistent calculation \cite{le1992self} in $D=2$ and $d=3$ dimensions are displayed in Table~\ref{tab:exponents}. With these scaling identites in hand, we can finally address the last key result of the paper, which is to show that the two ensembles are related to each other via a mechanical analog of Fisher renormalization \cite{fisher1968renormalization}. As the isotensional and isometric ensembles are thermodynamic duals of each other, the corresponding free energy densities are related to each other via a Legendre transformation (in the thermodynamic limit $V_D\to\infty$)
\begin{equation}
	F_\e=\min_{\sigma_0}\left(F_\sigma+\sigma_0\e\right)=F_\sigma(\sigma_*)+\sigma_*\e\;,\label{eq:legendre}
\end{equation}
where $\sigma_*$ solves $\e=-\partial F_\sigma/\partial\sigma_0|_{\sigma_0=\sigma_*}$, and the free energy densities are given by $F_{\sigma,\e}=-(k_BT/V_D)\ln\mathcal{Z}_{\sigma,\e}$. Now, near the buckling transition (at zero symmetry-breaking field $\mathcal{E}=0$), we can use the scaling theory developed above to obtain
\begin{equation}
	\Delta\e\sim|\sigma_*-\sigma_c|^{1/\theta}\implies\sigma_*-\sigma_c\sim|\Delta\e|^{\theta}\;,\label{eq:e*}
\end{equation}
where we have assumed $\theta>1$ and retained only the leading order term as $\Delta\e\to 0$.
Upon combining Eq.~\ref{eq:e*} with the scaling of the singular part of the free energy densities $F_\e\sim|\Delta\e|^{1+\theta'}$ and $F_\sigma\sim|\Delta\sigma|^{1+1/\theta}$ \footnote{The free energy densities also have nonsingular terms that need to be properly accounted for. In the limit $\Delta\sigma\to 0$ and $\Delta\e\to 0$, we have the following expansions for the respective free energy densities: $\mathcal{F}_\sigma=a_0+a_1\Delta\sigma+a_2|\Delta\sigma|^{1+1/\theta}+\mathcal{O}(\Delta\sigma^2)$ and $\mathcal{F}_\e=b_0+b_1\Delta\e+b_2|\Delta\e|^{1+\theta'}+\mathcal{O}(\Delta\e^2)$, where $a_{0,1,2}$ and $b_{0,1,2}$ are constants.}, we require that both sides of Eq.~\ref{eq:legendre} scale in the same way as $\Delta\sigma\to 0$. This constraint gives the equality
\begin{equation}
	\theta=\theta'\;.\label{eq:fisher}
\end{equation}
Although $\theta=\theta'$, from Eq.~\ref{eq:theta1} we immediately see that the correlation length exponents must differ in the two ensembles, $\nu\neq\nu'$. The simple form of the nontrivial relation in Eq.~\ref{eq:fisher} reflects the definition of $\theta,\theta'$ in Eq.~\ref{eq:theta}. Eqs.~\ref{eq:legendre},~\ref{eq:e*} now lead to
\begin{subequations}
\begin{gather}
	\beta'=\beta\theta\;,\quad\gamma'=\gamma\theta\;,\\
	\phi'=\phi\theta\;,\quad\nu'=\nu\theta\;.
\end{gather}\label{eq:fisher2}
\end{subequations}
The last of these relations can be solved using Eq.~\ref{eq:theta1} to explicitly give the important connection $\nu'=\nu/(\nu D-1)$.
With the help of Eq.~\ref{eq:nu2}, this relation simplifies to
\begin{equation}
	\nu'=\dfrac{1}{D-2+\eta}\;,
\end{equation}
which was obtained previously \cite{guitter1989thermodynamical}, without however recognizing a possible distinction in ensembles. In $D=2$, we get $\nu'=1/\eta$, which upon noting that $\eta<1$ leads to $\nu'>1$, which has been observed in old Monte-Carlo simulations of thermalized buckling in clamped sheets \cite{guitter1990stretching}. In general $D$, we can show that the isotensional and isometric ensembles have differing correlation length exponents such that
\begin{equation}
	\nu<\dfrac{2}{D}<\nu'\;.
\end{equation}
which is consistent with our renormalization group results in $D=4-\vare$ dimensions. Finally, demanding that Eq.~\ref{eq:fisher2} be consistent with Eqs.~\ref{eq:beta1},~\ref{eq:gamma1} leads to equality of the eta exponents in the two ensembles
\begin{equation}
	\eta'=\eta\;,\quad\eta'_u=\eta_u\;,\label{eq:etaequality}
\end{equation}
the latter being a consequence of the Ward identity (Eq.~\ref{eq:ward}).

We note that the difference in some of the exponents between the ensembles is not simply because the control variables ($\Delta\e$ and $\Delta\sigma$) have different dimensions. In fact, both the scaling variables we use have dimensions of stress and are equivalent to each other: we use $\Delta\sigma$ in the isotensional ensemble and $B\Delta\e$ (not $\Delta\e$) in the isometric ensemble. Hence the difference in exponents between the ensembles is due to a genuine change in the fixed point and its associated universality class, as confirmed by our renormalization group calculations.

These results are reminiscent of the Fisher renormalization of critical exponents due to hidden variables~\cite{fisher1968renormalization}, and are also related to the problems of a constrained \cite{rudnick1974renormalization} or a compressible magnets \cite{sak1974critical}, where the presence of a constraint (much like Eq.~\ref{eq:legendre}) leads to modified exponents. In conventional critical phenomena, such as in 3D magnets or superfluid He, Fisher renormalization doesn't affect the numerical values of exponents by much as it usually involves dividing the conventional exponents by $1-\alpha$, where $\alpha$ is the specific heat exponent, which is often a rather small correction \cite{fisher1968renormalization}. Here, however, the exponent $\theta$ replaces $1-\alpha$ in the mechanical context, allowing for a much stronger distinction in critical behaviour between the two ensembles.

In Table~\ref{tab:exponents}, we see that to leading order in an $\vare$-expansion, all the equalities in Eqs.~\ref{eq:fisher} and~\ref{eq:fisher2} are satisfied.
%The corresponding exponents evaluated in fixed dimension do numerically deviate by small amounts ($\sim10\%$) from the Fisher renormalization predictions, but that is to be expected as the one loop calculation for fixed $\{D,d\}$ is uncontrolled. The minor discrepencies suggest that higher loop corrections may not be very large.
Within the $\vare$-expansion, we find that the anomalous exponents are equal to leading order in the two ensembles, i.e., $\eta=\eta'$ and $\eta_u=\eta'_u$, as expected from our scaling considerations (Eq.~\ref{eq:etaequality}).
%The difference in numerical values between $\eta$ and $\eta'$ (correspondingly $\eta_u$ and $\eta'_u$) when computed at fixed dimension is presumably remedied when higher loop corrections are taken into account.
All the exponents in both ensembles are recapitulated in Table~\ref{tab:exponents} along with their exponent identities.

\section{Discussion}
\label{sec:discussion}
Although the study of thermalized membranes is more than three decades old \cite{nelson2004statistical}, it has been revitalized in recent years by enhanced interest in 2D materials such as graphene and MoS$_2$. Motivated by the ability to study extreme mechanics in such ultrathin materials \cite{Blees2015}, we have investigated the impact of thermal fluctuations on a classic (circa 1757!) Euler buckling instability of thin plates. By viewing the finite temperature buckling transition through the lens of critical phenomena, we have uncovered new exponent relations and remarkable phenomena that tie together geometry, mechanics and fluctuations in a thin elastic sheet. 

Near a thermodynamic continuous phase transition, fluctuations emerge on all scales and physics becomes universal. As we have shown, a similar situation arises on the verge of a mechanical instability, such as buckling, though with some surprises. The long-wavelength nature of the buckling transition leads to unusual critical scaling behaviour reflected in the system size dependence of the mechanical response. Additionally, buckling can be actuated under either isotensional (constant stress) or isometric (constant strain) loading, which as we have found, actually constitute separate universality classes. This remarkable feature highlights the importance of oft-neglected boundary conditions that when clamped can induce a novel thermally generated spontaneous tension and modify important scaling exponents. Our work demonstrates that the inequivalence of mechanical ensembles distinguished by their boundary conditions exemplifies the phenomenon of Fisher renormalization \cite{fisher1968renormalization} in a mechanical context.

We emphasize the salient role of geometry in isotropic thermalized buckling. Much of the phenomena discussed here arise due to the inevitable geometric coupling between in-plane stretching and out of plane bending, ubiquitous in thin plates but absent in lower dimensional counterparts, such as slender filaments. As a consequence, there is no analogue of our results in the finite temperature buckling of beams and polymers \cite{odijk1998microfibrillar,*hansen1999buckling,*baczynski2007stretching,*emanuel2007buckling,*bedi2015finite,*stuij2019stochastic}. Single-molecule measurements with polymers have noted an inequivalence of similar mechanical ensembles \cite{lubensky2002single,keller2003relating,*sinha2005inequivalence}, though in this case due to finite size effects. In contrast for thin sheets, the ensemble inequivalence at buckling survives the thermodynamic limit, as it instead originates from the tensionless flat phase being a critical phase, with fluctuations on all scales.

Our work is directly relevant to recent experiments that have probed the mechanics of 2D materials in various geometries \cite{Blees2015,nicholl2015effect,*nicholl2017hidden}. Boundary manipulation is a popular way to induce strong deformations and morphologies \cite{bao2009controlled,Vandeparre11}. When coupled to the material's electronic properties, this also allows for strain engineering synthetic gauge fields \cite{pereira2010geometry,*guinea2010energy}. Our results suggest that the interplay of thermal fluctuations and boundary constraints can be important in many of these contexts, allowing for enhanced control of the emergent mechanics in nanoscale systems. Anisotropic buckling is particularly relevant in such solid-state devices, either in terms of uniaxially compressed ribbons \cite{kovsmrlj2016response} or sheets crushed in the presence of a background aligning field \cite{leo2021}, both of which pose challenging directions for future research.
It is also known that the influence of boundaries can often persist in macroscopically large diffuse regions in slender elastic bodies \cite{schroll2011elastic}, an effect that is amplified by the geometry of plates \cite{lobkovsky1997properties,*cerda2004elements,*barois2014curved} and shells \cite{mahadevan2007persistence,*santangelo2013nambu}. It would be interesting to explore the consequence of thermal fluctuations in these cases, where boundary effects are again particularly important. 

While much of our analysis focused on the buckling transition, far beyond the threshold, the sheet adopts a curved profile whose mechanical description is akin to that of thin shells. The ensuing curvature can be manipulated to control localized deformations \cite{vaziri2008localized,*evans2017geometric} and fluctuation driven nonlinear response \cite{paulose2012fluctuating,*kovsmrlj2017statistical}. Such curved geometries offer tunable mechanisms to modulate the mechanical and vibrational properties of electromechanical resonators \cite{lifshitz2008nonlinear}, another attractive direction for future research. Postbuckled states also often exhibit bistability and sudden snap-through transitions that exhibit critical slowing down even in the absence of fluctuations \cite{gomez2017critical}. It would be of interest to extend our analysis to incorporate such hysteritic and dynamic effects at finite temperature, but this remains a formidable challenge. We hope that this work will encourage future explorations at the rich intersection of geometry, statistical mechanics and elastic instabilities.

\emph{Note added}: We recently became aware of related work by Leo Radzihovsky and Pierre Le Doussal \cite{leo2021} on the buckling transition in a thermalized membrane subjected to an external aligning field, which unlike our work, breaks rotational symmetry explicitly in the bulk.
\acknowledgments
SS acknowledges the support of the Harvard Society of Fellows. We thank Mark Bowick, Paul Hanakata, Andrej Ko{\v{s}}mrlj, John Toner and Leo Radzihovsky for illuminating discussions. This work was also supported by the NSF through the Harvard Materials Research Science and Engineering Center, via Grant No.~DMR-2011754 as well as by Grant No.~DMR-1608501.

\appendix
\section{Integrating out the in-plane phonons}
\label{app:iso}
As $\b{u}$ appears quadratically in the Hamiltonian (Eq.~\ref{eq:H}), we can exactly integrate it out. To do this, we separate the average strain ($u^0_{ij}$) from the nonzero wavelength deformations (i.e., $\b{q}\neq\b{0}$, denoted by the prime on the $\b{q}$ integral), to write
\begin{gather}
	u_{ij}(\b{r})=u_{ij}^0+\int_{\b{q}}'\;e^{i\b{q}\cdot\b{r}}\left[\dfrac{i}{2}(q_iu_j(\b{q})+q_ju_i(\b{q}))+\mathcal{A}_{ij}(\b{q})\right]\;,\\
	\textrm{with}\quad u^0_{ij}=\dfrac{1}{A}\int\dd^2r\;u_{ij}\;,\quad \mathcal{A}_{ij}(\b{r})=\dfrac{1}{2}\partial_ih\partial_jh\;,
\end{gather}
where $\int_{\b{q}}=\int\dd^2q/(2\pi)^2$. Note that, while there are only two independent in-plane phonon degrees of freedom ($u_i(\b{q})$) for nonzero wavevector, the homogeneous part of the strain tensor ($u^0_{ij}$) has \emph{three} independent components, corresponding to the three distinct modes of macroscopically deforming a 2D solid.
It is well known that only the transverse component of $\mathcal{A}_{ij}$ is important, as the rest can be absorbed into a global translation zero mode (constant displacement) \cite{nelson2004statistical}. The total Hamiltonian now takes the form $\mathcal{H}=\mathcal{H}'+\mathcal{H}^0$
\begin{widetext}
\begin{align}
	\mathcal{H}'&=\int\dfrac{\dd^2q}{(2\pi)^2}\left[\dfrac{\kappa}{2}q^4\left|h_\b{q}\right|^2+\dfrac{1}{2}u_i(\b{q})\left(\mu q^2\mathcal{P}^T_{ij}+(2\mu+\lambda)q^2\mathcal{P}^L_{ij}\right)u_j(-\b{q})\right]-\mathcal{E}h_{\b{q}=\b{0}}\nonumber\\
	&+\int'\dfrac{\dd^2q}{(2\pi)^2}\left[i\mu\left(q_iu_j(\b{q})+q_ju_i(-\b{q})\right)\mathcal{A}_{ij}(-\b{q})+i\lambda q_iu_i(\b{q})\mathcal{A}_{kk}(-\b{q})+\dfrac{1}{2}\left(2\mu|\mathcal{A}_{ij}(\b{q})|^2+\lambda|\mathcal{A}_{kk}(\b{q})|^2\right)\right]\;,\label{eq:H'}\\
	\mathcal{H}^0&=\dfrac{A}{2}\left[2\mu(u^0_{ij})^2+\lambda(u^0_{kk})^2\right]-A\sigma^{\rm ext}_{ij}\left(u^0_{ij}-\mathcal{A}^0_{ij}\right)\;,\label{eq:H0}
\end{align}
\end{widetext}
where $\mathcal{H}'$ includes all contributions from the $\b{q}\neq\b{0}$ in-plane phonon modes and $\mathcal{H}^0$ includes all the terms corresponding to the $\b{q}=\b{0}$ phonon modes.
Here we have used the transverse ($\mathcal{P}^T_{ij}(\b{q})=\delta_{ij}-q_iq_j/q^2$) and longitudinal ($\mathcal{P}^L(\b{q})=q_iq_j/q^2$) projection operators and written $\mathcal{A}^0_{ij}=(1/A)\int\dd\b{r}\;\mathcal{A}_{ij}(\b{r})$. The $\b{q}=\b{0}$ and $\b{q}\neq\b{0}$ in-plane phonon modes clearly decouple from each other, so when we integrate them out, the total free energy is simply $\mathcal{F}=\mathcal{F}'+\mathcal{F}^0$, where $\mathcal{F}'$ arises from integrating out the $\b{q}\neq\b{0}$ phonons and $\mathcal{F}^0$ arises from the $\b{q}=\b{0}$ phonon modes. The former is a standard calculation \cite{nelson2004statistical}, which gives
\begin{equation}
	\mathcal{F}'=\int\dd^2r\left\{\dfrac{\kappa}{2}(\del^2h)^2+\dfrac{Y}{2}\left(\dfrac{1}{2}\mathcal{P}^T_{ij}\partial_ih\partial_jh\right)^2-\mathcal{E}h\right\}\;.\label{eq:F'}
\end{equation}
Note that this part of the calculation is common to both ensembles. For the zero mode calculation, we consider the two ensembles separately. In the isotensional ensemble, $\sigma^{\rm ext}_{ij}=\sigma_0\delta_{ij}$, which gives
\begin{equation}
	\mathcal{H}^0_{\sigma}=\dfrac{A}{2}\left[2\mu(\tilde{u}^0_{ij})^2+(\mu+\lambda)(u^0_{kk})^2\right]-A\sigma_0\left(u^0_{kk}-\mathcal{A}^0_{kk}\right)\;,
\end{equation}
where we have decomposed $u^0_{ij}$ into its deviatoric part ($\tilde{u}^0_{ij}=u_{ij}^0-\delta_{ij}u_{kk}^0/2$) and its trace $u^0_{kk}$. Both the shear ($\tilde{u}_{ij}^0$) and the dilation ($u^0_{kk}$) components of the homogeneous strain can be integrated over freely now to obtain the zero mode contribution to the free energy,
\begin{equation}
	\mathcal{F}^0_{\sigma}=A\sigma_0\mathcal{A}^0_{kk}=\dfrac{\sigma_0}{2}\int\dd^2r\;|\vec{\del}h|^2\;,\label{eq:F0sigma}
\end{equation}
in the isotensional ensemble \cite{kovsmrlj2016response}. In the isometric ensemble, we set $\sigma^{\rm ext}_{ij}=0$ and instead have $(1/A)\int\dd\b{r}\vec{\del}\cdot\b{u}=\e$. The homogeneous part of the strain tensor is then given by
\begin{equation}
	u^0_{ij}=\tilde{u}^0_{ij}+\dfrac{\delta_{ij}}{2}\left(\e+\mathcal{A}^0_{kk}\right)\;,
\end{equation}
where we have once again separated out the devaitoric shear component ($\tilde{u}^0_{ij}$). We immediately see that, while in the isotensional ensemble, all three components of $u^0_{ij}$ were freely integrated over, in the isometric ensemble, only \emph{two} out of the three degrees of freedom can be freely integrated over. The clamped boundary conditions prevent homogeneous dilations or contractions, but the two homogeneous shear deformations in $\tilde{u}^0_{ij}$ continue to be zero modes. Upon integrating out $\tilde{u}^0_{ij}$, we obtain
\begin{align}
	\mathcal{F}^0_{\e}&=\dfrac{A}{2}(\mu+\lambda)\left(\e+\mathcal{A}^0_{kk}\right)^2\nonumber\\
	&=\dfrac{B}{2A}\int\dd^2r\left[\e+\dfrac{1}{2A}\int\dd^2r'|\vec{\del}'h|^2\right]^2\;,\label{eq:F0e}
\end{align}
with the bulk modulus $B=\mu+\lambda$, in the isometric ensemble. By adding together $\mathcal{F}'$ from Eq.~\ref{eq:F'} with $\mathcal{F}^0_{\sigma,\e}$ (Eqs.~\ref{eq:F0sigma} and~\ref{eq:F0e}), we get the total free energy in the two ensembles (Eqs.~\ref{eq:Fsigma} and~\ref{eq:Fe} in the main text).

\section{Mean field equation of state}
\label{app:mft}
For the mean field calculation we use a single mode Galerkin approximation using $h_0(\b{r})=H_0J_0(q_nr)$ for a general wavevector $q_n$. The linear terms are easily diagonalized by $h_0(\b{r})$ which is an eigenfunction of the Laplacian,
\begin{equation}
	\del^2h_0(\b{r})=-q_n^2h_0(\b{r})\;.
\end{equation}
The nonlinear terms are computed as follows. We first have the integral
\begin{equation}
	\int\dfrac{\dd^2r}{A}|\vec{\del}h_0|^2=H_0^2q_0^2f(q_0R)\;,
\end{equation}
where the dimensionless function is given by
\begin{align}
	f(x)&=\dfrac{2}{x^2}\int^x_0\dd r\; rJ_1(r)^2\nonumber\\
	&=J_0(x)^2+J_1(x)^2-\dfrac{2}{x}J_0(x)J_1(x)\;.
\end{align}
This is a rapidly oscillating function which vanishes at zero as $f(x)\simeq x^2/8$ ($x\to 0$) and has an envelope that asymptoticaly decays as $f(x)\sim 2/(\pi x)$ for $x\to\infty$. The second nonlinear term comes from nonlinear stretching and involves the projection operator which is easiest evaluated in Fourier space. Upon Fourier transforming, we have
\begin{equation}
	h_0(\b{q})=\int\dd^2r\;e^{-i\b{q}\cdot\b{r}}h_0(\b{r})=2\pi \dfrac{H_0}{q_n}\delta(q-q_n)\;,
\end{equation}
where $q=|\b{q}|$. We shall denote
\begin{equation}
	S(\b{r})=\dfrac{1}{2}\mathcal{P}_{ij}^T(\partial_jh_0\partial_jh_0)\;,
\end{equation}
and similarly define $S_{\b{q}}$ as its Fourier transform. This gives
\begin{align}
	S_{\b{q}}&=\int\dfrac{\dd^2k}{(2\pi)^2}\dfrac{1}{2}\mathcal{P}_{ij}^T(\b{q})k_ik_jh_0(\b{k})h_0(\b{q}-\b{k})\\
	&=\left(\dfrac{2\pi H_0}{q_n}\right)^2\int\dfrac{\dd^2 k}{(2\pi)^2}\dfrac{k^2}{2}\sin^2\vphi\delta(k-q_n)\delta(|\b{q}-\b{k}|-q_n)\;,
\end{align}
where $\cos\vphi=\hat{\b{q}}\cdot\hat{\b{k}}$. As the delta functions enforce $k=q_n$ and $\vphi=\pm\vphi_0=\pm\cos^{-1}(q/2k)$ ($|\sin\vphi_0|=\sqrt{1-(q/2k)^2}$), we can use the identity
\begin{equation}
	\delta(|\b{q}-\b{k}|-q_n)=\dfrac{q_n}{q k}\dfrac{\left[\delta(\vphi+\vphi_0)+\delta(\vphi-\vphi_0)\right]}{|\sin\vphi_0|}\;.
\end{equation}
After some simplifications, this then gives (with the restriction $q\leq 2q_n$)
\begin{equation}
	S(\b{q})=\dfrac{H_0^2}{2q}\sqrt{4q_n^2-q^2}\;.
\end{equation}
In order to project Eq.~\ref{eq:EL} onto $h_0(\b{r})$, we use the identity $\int_0^{\infty}\dd x\; J_0(x)=1$. This allows us to write
\begin{align}
	&q_n^2(\kappa q_n^2+\sigma)H_0+\dfrac{v}{2A}q_n^2H_0\int\dd^2r|\vec{\del}h_0(\b{r})|^2\nonumber\\
	&\quad-\dfrac{q_nY}{2\pi}\int\dd^2r\dfrac{1}{r}\mathcal{P}_{ij}^TS(\b{r})\partial_i\partial_jh_0(\b{r})=\dfrac{q_n\mathcal{E}}{2\pi}\int\dd^2r\dfrac{1}{r}\;.
0\end{align}
The divergent integral contribution from the external field is cutoff by the system size $R$ at large distances. The nonlinear stretching term can be equivalently computed in Fourier space by noting that
\begin{equation}
	\int\dd^2r\dfrac{F(\b{r})}{r}=\int\dfrac{\dd^2q}{(2\pi)^2}\dfrac{2\pi}{q}F_{\b{q}}\;,
\end{equation}
where $F_{\b{q}}$ is the Fourier transform of $F(\b{r})$.
Hence we compute this term in Fourier space as follows,
\begin{align}
	&q_n\int\dfrac{\dd q}{2\pi}\int\dd^2r\;e^{-i\b{q}\cdot\b{r}}\mathcal{P}_{ij}^TS(\b{r})\partial_i\partial_jh_0(\b{r})\nonumber\\
	&\ =-q_nH_0^3\int\dfrac{\dd q}{2\pi}\int\dfrac{\dd k}{2\pi}\dfrac{q}{k}\sqrt{4q_n^2-k^2}\sqrt{1-\dfrac{(q^2+k^2-q_n^2)}{4q^2k^2}}\nonumber\\
	&\ =-q_n^4H_0^3\int\dfrac{\dd x}{2\pi}\int\dfrac{\dd y}{2\pi}\dfrac{\sqrt{4-y^2}}{2y^2}\sqrt{4x^2y^2-(x^2+y^2-1)^2}\nonumber\\
	&\ =-c_0q_n^4H_0^3\;.
\end{align}
The integration domain for $x=q/q_n$ and $y=k/q_n$ is determined by the square root being real. This domain is a trapezoidal region in the first quadrant ($x,y\geq 0$) bounded by the lines $x+y=1$, $y=x+1$, $y=2$ and $y=x-1$. The integral evaluates to a constant $c_0\simeq 0.10567$. By combining the linear and nonlinear terms we then get the quoted Eq.~\ref{eq:eos}.

\begin{figure}[t]
	\centering
	\includegraphics[width=0.5\textwidth]{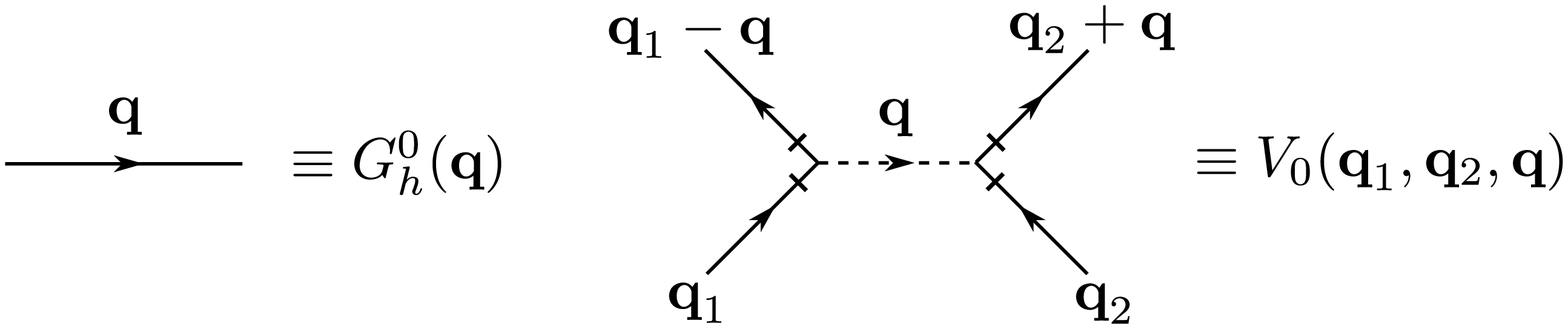}
	\caption{The bare propagator and four-point vertex for the height field, with the notation for labelling wave-vectors shown.}
	\label{fig:prop}
\end{figure}

\section{Renormalization group calculation: $\vare$-expansion}
\label{app:rgd}
Here we provide details of our renormalizaton group calculation in arbitrary dimensions. We consider a $D$-dimensional solid fluctuating in $d$-dimensional space, where $d_c=d-D>0$ is the codimension. Now, the in-plane displacement $\b{u}$ is a $D$-dimensional vector and the height field $\b{h}$ is a $d_c$-dimensional vector. The total elastic energy is once again of the form as in Eq.~\ref{eq:H}, though with the nonlinear strain tensor
\begin{equation}
	u_{ij}=\dfrac{1}{2}\left(\partial_iu_j+\partial_ju_i+\partial_i\b{h}\cdot\partial_j\b{h}\right)\;.
\end{equation}
We follow the standard calculation of integrating out the phonons to now obtain an effective free energy solely as a function of the height $\b{h}$ \cite{le1992self,aronovitz1988fluctuations,guitter1988crumpling}. We handle the boundary conditions in the two ensembles as detailed in Appendix~\ref{app:iso} to obtain
\begin{align}
	\mathcal{F}&=\dfrac{1}{2}\int\dd^Dr\left[\kappa|\del^2\b{h}|^2+\sigma|\vec{\del}\b{h}|^2\right]\nonumber\\
	&\quad+\dfrac{1}{4}\int\dd^Dr(\partial_i\b{h}\cdot\partial_j\b{h})R_{ijk\ell}(\partial_k\b{h}\cdot\partial_{\ell}\b{h})\nonumber\\
	&\quad+\dfrac{v}{8V_D}\int\dd^Dr\int\dd^Dr'|\vec{\del}\b{h}|^2|\vec{\del}'\b{h}|^2\;,
\end{align}
where $v$, as before, distinguishes the two ensembles and $V_D$ is the volume of the $D$-dimensional solid.
This free energy extends Eq.~\ref{eq:F} to general dimensions. We have set the external field to zero ($\vec{\mathcal{E}}=\b{0}$, which now has $d_c$ components) as it won't be important for the diagrammatic calculation. The isotensional ensemble corresponds to $\sigma=\sigma_0$ (the external isotropic stress) and $v=0$. The isometric ensemble corresponds to setting $\sigma=B\e$ and $v=B$, where $B=(2\mu/D)+\lambda$ is the $D$-dimensional generalization of the bulk modulus and $\e$ is the external isotropic strain imposed. The nonlinear stretching term is given by $R_{ijk\ell}(\b{q})=\mu M_{ijk\ell}(\b{q})+(Y/2)N_{ijk\ell}(\b{q})$ \cite{le1992self}, where $Y=2\mu(2\mu+D\lambda)/(2\mu+\lambda)$ is the $D$-dimensional version of the Young's modulus and
\begin{align}
	N_{ijk\ell}(\b{q})&=\dfrac{1}{D-1}\mathcal{P}_{ij}^T(\b{q})\mathcal{P}_{k\ell}^T(\b{q})\;\\
	M_{ijk\ell}(\b{q})&=\dfrac{1}{2}\left[\mathcal{P}_{ik}^T(\b{q})\mathcal{P}_{j\ell}^T(\b{q})+\mathcal{P}_{i\ell}^T(\b{q})\mathcal{P}_{jk}^T(\b{q})\right]-N_{ijk\ell}(\b{q})\;.
\end{align}
In $D=2$, $M_{ijk\ell}$ vanishes identically. This decomposition is useful as $\b{M}$ and $\b{N}$ are mutually orthogonal tensors. We can compute the one-loop correction to the bending rigidity and the elastic moduli using this free energy. Note that, unlike in the $D=2$ case, both $\mu$ and $Y$ appear separately in the reduced free energy for $D>2$. The bare propagator (correlator) for the height field is defined via $\langle h_\mu(\b{q})h_\nu(-\b{q})\rangle_0/V_D=\delta_{\mu\nu}G^0_h(\b{q})$, where $\mu,\nu=1,\cdots,\,d_c$, and its renormalized version is $G_h(\b{q})$.
As both the nonlinearities arising from $\b{R}$ and $v$ are quartic in nature, we combine the two into a single interaction vertex for simplicity,
\begin{gather}
	\mathcal{F}_{\rm int}=\int_{\b{q}}\int_{\b{q}_1}\int_{\b{q}_2}V_0(\b{q}_1,\b{q}_2,\b{q})(\b{h}_{\b{q}_1}\cdot\b{h}_{\b{q}-\b{q}_1})(\b{h}_{\b{q}_2}\cdot\b{h}_{-\b{q}-\b{q}_2})\;,\\
	V_0(\b{q}_1,\b{q}_2,\b{q})=\dfrac{1}{4}R_{ijk\ell}(\b{q})q_{1i}q_{1j}q_{2k}q_{2\ell}\nonumber\\
	\quad+\dfrac{v}{8V_D}q_1^2q_2^2\;(2\pi)^D\delta(\b{q})\;,\label{eq:Vbare}
\end{gather}
where $\int_{\b{q}}=\int\dd^Dq/(2\pi)^D$.
Note that the usual nonlinear stretching term excludes the zero mode \cite{nelson1987fluctuations,nelson2004statistical,le1992self}. In other words, we always work in the convention that $\mathcal{P}_{ij}^T(\b{q}=\b{0})=0$, hence $R_{ijk\ell}(\b{q}=\b{0})=0$. Both $G_h^0(\b{q})$ and $V_0(\b{q}_1,\b{q}_2,\b{q})$ are graphically represented in Fig.~\ref{fig:prop}. The new nonlocal nonlinear term $\propto v/8V_D$ is unusual as it is ultra-local in Fourier space, with a delta function in $\b{q}$. After all, the new nonlinear term arose from integrating out the strain zero mode in the isometric ensemble, hence it makes sense that the associated nonlinearity has strict support on $\b{q}=\b{0}$. In conjunction with the fact that $\mathcal{P}_{ij}^T(\b{0})=0$, the nonlinear couplings are orthogonal to each other and their associated operators don't mix.

\begin{figure*}[]
	\centering
	\includegraphics[width=\textwidth]{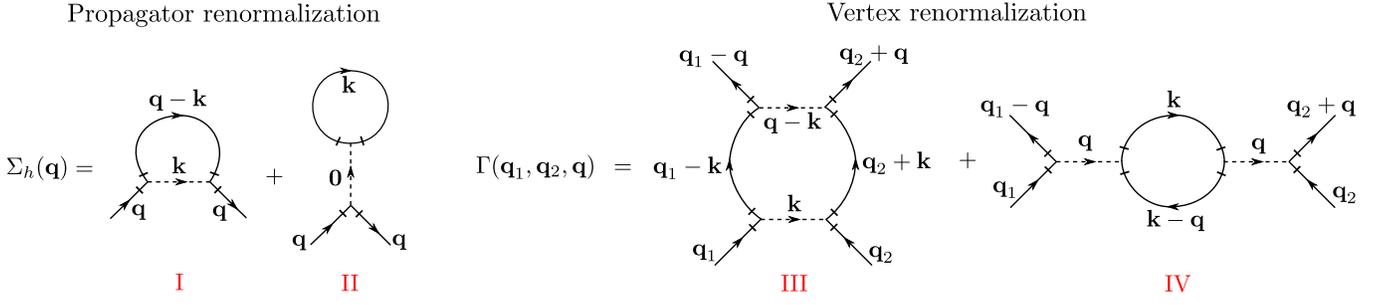}
	\caption{One-loop contribution to both the self-energy $\Sigma_h(\b{q})$ and the vertex correction $\Gamma(\b{q}_1,\b{q}_2,\b{q})$.}
	\label{fig:rg}
\end{figure*}
We perform standard Wilsonian renormalization \cite{goldenfeld2018lectures} by integrating out a shell of wavevectors $\Lambda/b\leq q\leq\Lambda$, where $\Lambda$ is the UV cutoff. The renormalized propagator is given by Dyson's equation,
\begin{equation}
	G_{h}^{-1}(\b{q})=G_h^0(\b{q})^{-1}-\Sigma_h(\b{q})\;,
\end{equation}
where $\Sigma_h(\b{q})$ is the ``self-energy'' and the vertex corrections are encapsulated in $\Gamma(\b{q}_1,\b{q}_2,\b{q})$ with $\beta V=\beta V_0-\Gamma$ ($\beta=1/k_BT$, not to be confused with the order parameter exponent). Upon including the relevant combinatorial factors, we obtain at one-loop order
\begin{widetext}
\begin{align}
	\Sigma_h(\b{q})&=8(-\beta)\times\int\dfrac{\dd^Dk}{(2\pi)^D}V_0(\b{q},-\b{q},\b{k})G_h^0(\b{q}-\b{k})+4d_c(-\beta)\times\int\dfrac{\dd^Dk}{(2\pi)^D}V_0(\b{q},\b{k},\b{0})G_h^0(\b{k})\;,\\
	\Gamma(\b{q}_1,\b{q}_2,\b{q})&=32\dfrac{(-\beta)^2}{2!}\times\int\dfrac{\dd^Dk}{(2\pi)^D}V_0(\b{q}_1,\b{q}_2,\b{q})G_h^0(\b{q}_1-\b{k})G_h^0(\b{q}_2+\b{k})V_0(\b{q}_1-\b{k},\b{q}_2+\b{k},\b{q}-\b{k})\nonumber\\
	&\quad+8d_c\dfrac{(-\beta)^2}{2!}\times\int\dfrac{\dd^Dk}{(2\pi)^D}V_0(\b{q}_1,\b{k}-\b{q},\b{q})G_h^0(\b{k})G^0_h(\b{k}-\b{q})V_0(\b{q}-\b{k},\b{q}_2,\b{q})\;.
\end{align}
\end{widetext}
We can expand these out for small wavevectors ($q\to 0$) and perform the $\b{k}$ integrals over a thin slice $\Lambda/b<k<\Lambda$ by leverage standard results \cite{le1992self} for the loop integrals. To handle the new nonlinear coupling $v$, we use the fact that $\mathcal{P}_{ij}^T(\b{0})=0$ to set $\delta(\b{q})\mathcal{P}_{ij}^T(\b{q})=0$. As a result diagram III in Fig.~\ref{fig:rg} does not contribute to the vertex correction at long wavelengths. Both $\Sigma_h$ and $\Gamma$ then have singular terms involving the delta function, which we regularize as,
\begin{subequations}
\begin{gather}
	(2\pi)^D\delta(\b{0})=\int\dd^Dr\equiv V_D\;,\\
	\left[(2\pi)^D\delta(\b{q})\right]^2=(2\pi)^D\delta(\b{0})(2\pi)^D\delta(\b{q})\equiv V_D(2\pi)^D\delta(\b{q})\;.
\end{gather}
	\label{eq:reg}
\end{subequations}
This cancels the extra factors of $V_D$ leaving all the coupling constants as intensive parameters, as expected. We cast the perturbative correction for the various coupling constants into differential recursion relations by writing $\ln b= s\ll 1$ and implementing the scaling transformation to restore the UV cutoff, which gives
\begin{widetext}
\begin{subequations}
\begin{align}
	\dfrac{\dd\kappa}{\dd s}&=(2\zeta-4+D)\kappa+\dfrac{2k_BT(D+1)}{D(D+2)}\dfrac{\Lambda^{D-2}S_{D-1}}{(2\pi)^D(\kappa\Lambda^2+\sigma)}\left[\dfrac{Y}{2}+(D-2)\mu\right]\;,\\
	\dfrac{\dd\sigma}{\dd s}&=(2\zeta-2+D)\sigma+\dfrac{k_BTd_c}{2}\dfrac{v\Lambda^D S_{D-1}}{(2\pi)^D(\kappa\Lambda^2+\sigma)}\;,\\
	\dfrac{\dd v}{\dd s}&=(4\zeta-4+D)v-\dfrac{k_BTd_c}{2}\dfrac{v^2\Lambda^{D}S_{D-1}}{(2\pi)^D(\kappa\Lambda^2+\sigma)^2}\;,\\
	\dfrac{\dd \mu}{\dd s}&=(4\zeta-4+D)\mu-\dfrac{2k_BTd_c}{D(D+2)}\dfrac{\mu^2\Lambda^DS_{D-1}}{(2\pi)^D(\kappa\Lambda^2+\sigma)^2}\;,\\
	\dfrac{\dd Y}{\dd s}&=(4\zeta-4+D)Y-\dfrac{k_BTd_c(D+1)}{2D(D+2)}\dfrac{Y^2\Lambda^DS_{D-1}}{(2\pi)^D(\kappa\Lambda^2+\sigma)^2}\;,\\
	\dfrac{\dd B}{\dd s}&=(4\zeta-4+D)B-\dfrac{k_BTd_c}{2}\dfrac{B^2\Lambda^{D}S_{D-1}}{(2\pi)^D(\kappa\Lambda^2+\sigma)^2}\;,\\
	\dfrac{\dd\nu_p}{\dd s}&=-\dfrac{k_BTd_c}{D(D+2)}\dfrac{\mu\Lambda^D S_{D-1}}{(2\pi)^D(\kappa\Lambda^2+\sigma)^2}(1+\nu_p)(1+3\nu_p)\;,
\end{align}\label{eq:SIrgd}
\end{subequations}
\end{widetext}
where $S_{D-1}=2\pi^{D/2}/\Gamma(D/2)$ is the volume of a unit sphere in $D$-dimensions.
These generalize the recursion relations in the main text (Eqs.~\ref{eq:dkappadl}-\ref{eq:dnudl}) to arbitrary $D$ and $d_c$, which are recovered by simply setting $d_c=1$and $D=4-\vare$, and retaining terms only to $\mathcal{O}(\vare)$. Note that both $Y$ and $\mu$ enter separately to renormalize $\kappa$ when $D>2$. As expected, the shear modulus $\mu$, Young's modulus $Y$ and bulk modulus $B$, all renormalize independently. Of course, while we provide the recursion relations for $\mu,Y,B$ and $\nu_p$ they are all related, and only two are independent. Furthermore, $B$ and $v$ renormalize in identical ways as is required to be consistent with the isometric ensemble.

In order to analyze these equations, we once again switch to dimensionless variables just as before (now in $D$-dimensions)
\begin{subequations}
\begin{gather}
	K=\dfrac{\kappa\Lambda^2}{\kappa\Lambda^2+\sigma}\;,\quad\bar{Y}=\dfrac{k_BTY\Lambda^D}{(\kappa\Lambda^2+\sigma)^2}\;,\\
	\bar{v}=\dfrac{k_BTv\Lambda^D}{(\kappa\Lambda^2+\sigma)^2}\;,\quad\bar{\mu}=\dfrac{k_BT\mu\Lambda^D}{(\kappa\Lambda^2+\sigma)^2}\;.
\end{gather}\label{eq:SIdimless}
\end{subequations}
The recursion relations for these dimensionless variables can be obtained after a fair bit of tedious algebra but we do not quote them here as the equations are cumbersome and not very illuminating. Instead we directly proceed to the fixed points. As $D=4$ is the upper critical dimension \cite{aronovitz1988fluctuations,guitter1988crumpling}, we set $D=4-\vare$ and work within an $\vare$-expansion. We have two interacting fixed points, one with $\bar{v}=0$ (vK$_{\rm th}$) appropriate for the isotensional ensemble and another with $\bar{v}\neq 0$ (CvK$_{\rm th}$) associated with the isometric ensemble. To leading order in $\vare$ and arbitrary $d_c$, the fixed points are given by
\begin{align}
	&{\rm vK_{th} }:\nonumber\\
	&K_*=1\;,\ \bar{Y}_*=\dfrac{385\pi^2\vare}{5(24+d_c)}\;,\ \bar{\mu}_*=\dfrac{96\pi^2\vare}{(24+d_c)}\;,\ \bar{v}_*=0\;.\\
	&{\rm CvK_{th} }:\nonumber\\
	&K_*=1+\dfrac{d_c\vare}{(48+2d_c)}\;,\ \bar{Y}_*=\dfrac{385\pi^2\vare}{5(24+d_c)}\;,\nonumber\\
	&\bar{\mu}_*=\dfrac{96\pi^2\vare}{(24+d_c)}\;,\ \bar{v}_*=\dfrac{16\pi^2\vare}{(24+d_c)}\;.
\end{align}
As before, the constrained fixed point CvK$_{\rm th}$ involves bare compression (as $K_*>1$) signalling the presence of a spontaneous thermal tension that is absent in the unconstrained fixed point vK$_{\rm th}$. One can also check that, in both ensembles, the $D$-dimensional version of the Poisson's ratio flows to its stable attracting fixed point given by
\begin{equation}
	\nu_p=\dfrac{\lambda}{2\mu+(D-1)\lambda}=-\dfrac{1}{3}\;,
\end{equation}
independent of both $D$ (conversely $\vare$) and $d_c$. This confirms the universal Poisson's ratio obtained through more sophisticated self-consistent calculations as well \cite{le1992self}.

Now, we can linearize about these fixed points to obtain the relevant anomalous scaling dimensions. For the isotensional ensemble, we fix $\bar{v}=0$ and diagonalize the Jacobian matrix about vK$_{\rm th}$ to obtain the eigenvalues
\begin{equation}
	y_0=2-\dfrac{12\vare}{(24+d_c)}\;,\ y_1=-\dfrac{d_c\vare}{(24+d_c)}\;,\ y_2=-\vare\;.
\end{equation}
As expected, we have two irrelevant directions ($y_{1,2}$) corresponding to $\bar{Y}$ and $\bar{\mu}$ and one unstable or relevant direction ($y_0$) corresponding to $K$. This provides the correlation length exponent as $\nu=1/y_0$, which is quoted in Table~\ref{tab:exponents}. The anomalous exponent $\eta$ is obtained by tuning right to the fixed point and evaluating
\begin{equation}
	\eta=\left.\dfrac{(D+1)S_{D-1}[\bar{Y}_*+2(D-2)\bar{\mu}_*]}{(2\pi)^DD(2+D)K_*}\right|_{\rm vK_{th}}=\dfrac{12\vare}{(24+d_c)}\;,
\end{equation}
to first order in $\vare$. As can be checked, this also satisfies the relation $\nu^{-1}=2-\eta$ (Eq.~\ref{eq:nu2}). The rest of the exponents quoted in Table~\ref{tab:exponents} are obtained by using the various exponent relations derived in the main text.

Similarly, now allowing for $\bar{v}\neq 0$ in the isometric ensemble, we can diagonalize the Jacobian matrix about CvK$_{\rm th}$ to obtain the eigenvalues
\begin{equation}
	y_0=2-\dfrac{(12+d_c)\vare}{(24+d_c)}\;,\ y_1=y_2=-\dfrac{d_c\vare}{(24+d_c)}\;,\ y_3=-\vare\;.
\end{equation}
We have three irrelevant directions ($y_{1,2,3}$) corresponding to $\bar{Y}$, $\bar{\mu}$ and $\bar{v}$, and one relevant direction along $K$. The correlation length exponent is obtained via $\nu'=1/y_0$, which is given in Table~\ref{tab:exponents}. The anomalous dimension $\eta'$ is computed just as in the isotensional case
\begin{equation}
	\eta'=\left.\dfrac{(D+1)S_{D-1}[\bar{Y}_*+2(D-2)\bar{\mu}_*]}{(2\pi)^DD(2+D)K_*}\right|_{\rm CvK_{th}}=\dfrac{12\vare}{(24+d_c)}\;.
\end{equation}
To leading order in $\vare$, we find $\eta=\eta'$ consistent with the scaling identity in Eq.~\ref{eq:etaequality}. The other exponents in the isometric ensemble are computed through the various exponent identites and are reported in Table~\ref{tab:exponents}.

\section{Renormalization group calculation: fixed dimension}
\label{app:rg}
Here we provide the details for deriving the recursion relations by using an uncontrolled one-loop approximation at fixed dimension ($D=2,d=3$). As $d_c=1$ here, the height field is a simple scalar, while the in-plane phonons are 2D vectors. The bare propagators (correlators) for the height and phonon fields are respectively $G_h^0(\b{q})$ and $\b{G}_u^0$, while their renormalized versions are denoted by $G_h(\b{q})$ and $\b{G}_u(\b{q})$. The bare quartic interaction vertex is now written as
\begin{gather}
	\mathcal{F}_{\rm int}=\int_{\b{q}}\int_{\b{q}_1}\int_{\b{q}_2}V_0(\b{q}_1,\b{q}_2,\b{q})h_{\b{q}_1}h_{\b{q}_2}h_{\b{q}-\b{q}_1}h_{-\b{q}-\b{q}_2}\;,\\
	V_0(\b{q}_1,\b{q}_2,\b{q})=\dfrac{Y}{8}[\b{q}_1\cdot\mathcal{P}^T(\b{q})\cdot\b{q}_1][\b{q}_2\cdot\mathcal{P}^T(\b{q})\cdot\b{q}_2]\nonumber\\
	\quad+\dfrac{v}{8A}q_1^2q_2^2\;(2\pi)^2\delta(\b{q})\;,\label{eq:Vbare2}
\end{gather}
where $\int_{\b{q}}=\int\dd^2q/(2\pi)^2$. As expected, in 2D, only the Young's modulus $Y$ enters the nonlinear stretching term, which, once again, excludes the zero mode \cite{nelson1987fluctuations,nelson2004statistical} ($\mathcal{P}_{ij}^T(\b{q}=\b{0})=0$).

We once again perform standard Wilsonian renormalization \cite{goldenfeld2018lectures} by integrating out a shell of wavevectors $\Lambda/b\leq q\leq\Lambda$, where $\Lambda$ is the UV cutoff. The self energy and vertex corrections are defined as before and computed to one-loop order.
Upon expanding them for small wavevectors ($q\to 0$) we obtain,
\begin{gather}
	\Sigma_h(\b{q})=-\dfrac{v q^2}{4\pi}\dfrac{\Lambda^2\ln b}{(\kappa\Lambda^2+\sigma)}-\dfrac{3Y}{16\pi}\dfrac{q^4\ln b}{(\kappa\Lambda^2+\sigma)}\;,\\
	\Gamma(\b{q}_1,\b{q}_2,\b{q})=\mathcal{P}_{ij}^T(\b{q})\mathcal{P}_{k\ell}^T(\b{q})q_{1i}q_{1j}q_{2k}q_{2\ell}\dfrac{Y^2}{8}\dfrac{3\Lambda^2\ln b}{32\pi(\kappa\Lambda^2+\sigma)^2}\nonumber\\
	\quad+q_1^2q_2^2\dfrac{v^2}{8A}(2\pi)^2\delta(\b{q})\dfrac{\Lambda^2\ln b}{4\pi(\kappa\Lambda^2+\sigma)^2}\;.
\end{gather}
We have once again used the fact that $\mathcal{P}_{ij}^T(\b{0})=0$ to set $\delta(\b{q})\mathcal{P}_{ij}^T(\b{q})=0$ along with the regularization scheme in Eq.~\ref{eq:reg} to eliminate factors of $A$.
By writing $\ln b= s\ll 1$ and implementing the scaling transformation to restore the UV cutoff, we obtain the following differential recursion relations
\begin{subequations}
\begin{align}
	\dfrac{\dd\kappa}{\dd\ell}&=\kappa(2\zeta-2)+k_BT\dfrac{3Y}{16\pi(\kappa\Lambda^2+\sigma)}\;,\\
	\dfrac{\dd\sigma}{\dd\ell}&=\sigma2\zeta+k_BT\dfrac{v\Lambda^2}{4\pi(\kappa\Lambda^2+\sigma)}\;,\\
	\dfrac{\dd Y}{\dd\ell}&=Y(4\zeta-2)-k_BT\dfrac{3Y^2\Lambda^2}{32\pi(\kappa\Lambda^2+\sigma)^2}\;,\\
	\dfrac{\dd v}{\dd\ell}&=v(4\zeta-2)-k_BT\dfrac{v^2\Lambda^2}{4\pi(\kappa\Lambda^2+\sigma)^2}\;,
\end{align}
\end{subequations}
which match Eqs.~\ref{eq:SIrgd} upon setting $D=2$ and $d_c=1$. In the $v=0$ limit, we also recover the fixed dimension recursion relations derived previously in Ref.~\cite{kovsmrlj2016response}.

We can also independently compute the renormalization of the elastic moduli from the fluctuation correction to the phonon propagator. Looking back at the full Hamiltonian in Eq.~\ref{eq:H}, we can similarly define the renormalized phonon propogator and associated self-energy via
\begin{equation}
	[\b{G}_u(\b{q})]^{-1}=[\b{G}_u^0(\b{q})]^{-1}-\vec{\Sigma}_u\;.
\end{equation}
At one-loop order, the phonon self energy is given by
\begin{widetext}
\begin{equation}
	\left[\vec{\Sigma}_u(\b{q})\right]_{ij}=2\times\dfrac{(-\beta)^2}{2!}\times2\int\dfrac{\dd^2k}{(2\pi)^2}U^0_i\left(\b{q},\dfrac{\b{q}}{2}+\b{k}\right)U^0_j\left(-\b{q},-\dfrac{\b{q}}{2}-\b{k}\right)G^0_h\left(\dfrac{\b{q}}{2}+\b{k}\right)G^0_h\left(\dfrac{\b{q}}{2}-\b{k}\right)\;,
\end{equation}
\end{widetext}
where $U^0_i(\b{q},\b{q}_1)=(-i/2)\{\lambda q_i[\b{q}_1\cdot(\b{q}-\b{q}_1)]+\mu[(\b{q}\cdot\b{q}_1)(q_i-q_{1i})+(\b{q}\cdot(\b{q}-\b{q}_1))q_{1i}]\}$ is the bare phonon-height cubic interaction vertex.
Upon expanding this self energy for small $q$ gives the following corrections to the Lam{\'e} coefficients,
\begin{gather}
	\mu'=\mu-\dfrac{k_BT\mu^2\Lambda^2\ln b}{8\pi(\kappa\Lambda^2+\sigma)^2}\;,\\
	(2\mu'+\lambda')=(2\mu+\lambda)-\dfrac{k_BT\Lambda^2\ln b}{4\pi(\kappa\Lambda^2+\sigma)^2}\left[(\mu+\lambda)^2+\dfrac{\mu^2}{2}\right]\;.
\end{gather}
This too can be cast into differential recursion relations for the shear modulus ($\mu$), the bulk modulus ($B=\mu+\lambda$) as well as the Poisson's ratio ($\nu_p=\lambda/(2\mu+\lambda)$), which gives
\begin{subequations}
\begin{align}
	\dfrac{\dd\mu}{\dd\ell}&=\mu(4\zeta-2)-k_BT\dfrac{\mu^2\Lambda^2}{8\pi(\kappa\Lambda^2+\sigma)^2}\;,\\
	\dfrac{\dd B}{\dd\ell}&=B(4\zeta-2)-k_BT\dfrac{B^2\Lambda^2}{4\pi(\kappa\Lambda^2+\sigma)^2}\;,\\
	\dfrac{\dd\nu}{\dd\ell}&=-k_BT\dfrac{\mu\Lambda^2}{16\pi(\kappa\Lambda^2+\sigma)^2}(1+\nu)(1+3\nu)\;,
\end{align}
\end{subequations}
once again consistent with Eq.~\ref{eq:SIrgd} upon setting $D=2$ and $d_c=1$.

While the one-loop approximation is uncontrolled, we can nonetheless obtain the fixed points and scaling exponents within this approximation. To do so we now use the 2D versions of the dimensionless variables (set $D=2$ in Eq.~\ref{eq:SIdimless}) to get the following recursion relations
\begin{subequations}
\begin{align}
	\dfrac{\dd K}{\dd\ell}&=2(K-1)\left[K-\dfrac{3\bar{Y}}{32\pi}\right]-\dfrac{\bar{v}K}{4\pi}\;,\\
	\dfrac{\dd\bar{Y}}{\dd\ell}&=\left[4K-2-\dfrac{\bar{v}}{2\pi}-\dfrac{15\bar{Y}}{32\pi}\right]\bar{Y}\;,\\
	\dfrac{\dd\bar{B}}{\dd\ell}&=\left[4K-2-\dfrac{\bar{v}}{2\pi}-\dfrac{\bar{B}}{4\pi}-\dfrac{3\bar{Y}}{8\pi}\right]\bar{B}\;.\\
	\dfrac{\dd\bar{v}}{\dd\ell}&=\left[4K-2-\dfrac{3\bar{v}}{4\pi}-\dfrac{3\bar{Y}}{8\pi}\right]\bar{v}\;.\\
	\dfrac{\dd\nu}{\dd\ell}&=-\dfrac{\bar{Y}}{32\pi}(1+3\nu)\;.
\end{align}
\end{subequations}
These can be solved to obtain two physically relevant interacting fixed points given by
\begin{itemize}
	\item vK$_{\rm{th}}$: $K_*=1$, $\bar{Y}_*=64\pi/15$, $\bar{B}_*=8\pi/5$, $\bar{v}_*=0$ ($\nu_p=-1/3$).
	\item CvK$_{\rm{th}}$: $K_*=\sqrt{2}$, $\bar{Y}_*=64\pi(2\sqrt{2}-1)/21$, $\bar{B}_*=8\pi(2\sqrt{2}-1)/7$, $\bar{v}_*=8\pi(2\sqrt{2}-1)/7$ ($\nu_p=-1/3$).
\end{itemize}
As before, the vK$_{\rm{th}}$ fixed point controls buckling in the isotensional ensemble and it matches previous fixed dimension calculations \cite{kovsmrlj2016response}, while CvK$_{\rm{th}}$ is new and controls buckling in the isometric ensemble. The universal Poisson's ratio in both ensembles is once again $\nu_p=-1/3$.

In the isotensional ensemble, we set $\bar{v}=0$ and linearize about vK$_{\rm{th}}$ to obtain the Jacobian eigenvalues
\begin{equation}
	y_0=\dfrac{6}{5}\;,\ y_1=-2\;,\ y_3=-\dfrac{2}{5}\;,
\end{equation}
with two irrelevant directions along $\bar{Y}$ and $\bar{B}$ and one relevant direction along $K$. This directly gives the correlation length exponent as $\nu=1/y_0=5/6$. The anomalous dimension is given by
\begin{equation}
	\eta=\left.\dfrac{3\bar{Y}_*}{16\pi K_*}\right|_{\rm{vK_{th}}}=\dfrac{4}{5}\;,
\end{equation}
matching the value obtained in Ref.~\cite{kovsmrlj2016response}.
Both the exponents ($\nu$ and $\eta$), computed within this uncontrolled approximation are fortuitously close in numerical value to more accurate estimates of the exponents produced via self-consistent calculations \cite{le1992self} (see also Table~\ref{tab:exponents}).

In the isometric ensemble, we now allow $\bar{v}>0$ and linearize about the new fixed point CvK$_{\rm{th}}$. The eigenvalues of the corresponding Jacobian matrix are irrational and given by
\begin{equation}
	y_0=1.1056\;,\ y_1=-2.6729\;,\ y_2=y_3=-0.5224\;,
\end{equation}
with three irrelevant directions ($\bar{Y},\bar{B}$ and $\bar{v}$) and one relevant direction ($K$). We once again obtain the correlation length exponent simply via $\nu'=1/y_0\approx0.9045$. This doesn't satisfy the inequality $\nu'>1$, nor does it satisfy $\nu'=1/\eta$ (with $\eta=4/5$ from above), both of which are expected in $D=2$ from general scaling arguments (Sec.~\ref{sec:scaling}). The discrepancy is attributed to the uncontrolled nature of the one-loop approximation here.
We can similarly compute the anomalous scaling dimension at this fixed point, to get
\begin{equation}
	\eta'=\left.\dfrac{3\bar{Y}_*}{16\pi K_*}\right|_{\rm{CvK_{th}}}=\dfrac{2}{7}(4-\sqrt{2})\approx0.7388\;,
\end{equation}
which evidently does not satisfy the identity $\eta=\eta'$ (Eq.~\ref{eq:etaequality}). Once again, we attribute this discrepancy to the uncontrolled one-loop approximation.

\section{Simplified scaling form \& Widom's identity}
\label{app:simple}
Here we formulate a scaling form for the order parameter $\langle h\rangle$ that is a function of a single (albeit size dependent) scaling variable.
A comparison of the size-dependent asymptotics in Eqs.~\ref{eq:psi0},~\ref{eq:psi10} and~\ref{eq:psiinf} suggests such a simpler scaling solution is possible. This allows us to write
\begin{equation}
	\Psi_h\left(\dfrac{\mathcal{E}}{|\Delta\sigma|^{\phi}}\right)=\dfrac{R}{Y^{1/2}}|\Delta\sigma|^{\nu(1+\eta_u/2)}\Phi_h\left(\dfrac{\mathcal{E}RY^{1/2}}{|\Delta\sigma|^f}\right)\;,
\end{equation}
that is now solely a function of one scaled variable with \emph{no} additional dependence on system size or elastic moduli. This form behaves correctly near zero, provided $\Phi_h(0)$ and $\Phi_h'(0)$ both approach finite constants. We can now ensure the correct asymptotics (and now $R,Y$ independent) by demanding that $\Phi_h(x)\sim x^{1/\delta}$ for $x\to\infty$. The modified gap exponent $f=\phi-\nu(1-\eta_u/2)=3\nu(2-\eta)/2$. This observation leads to an alternate scaling form for $\langle h\rangle$, namely
\begin{equation}
	\langle h\rangle=|\Delta\sigma|^{\beta}\dfrac{R}{Y^{1/2}}\Phi_h\left(\dfrac{\mathcal{E}RY^{1/2}}{|\Delta\sigma|^f}\right)\;,
\end{equation}
from which we can easily derive the exponent relations,
\begin{gather}
	f=\beta+\gamma\;,\quad f'=\beta'+\gamma'\;,\\
	\delta\beta=\gamma+\beta\;,\quad\delta'\beta'=\gamma'+\beta'\;.\label{eq:widom}
\end{gather}
Eq.~\ref{eq:widom} is Widom's identity \cite{goldenfeld2018lectures}. Both these equations are consistent with our previous results given in Eqs.~\ref{eq:beta1},~\ref{eq:gamma1},~\ref{eq:g2b} and~\ref{eq:delta1}. The modified gap exponent $f=3/2$, in the isotensional ensemble, which is exact and independent of dimension. The modified gap exponent in the isometric ensemble takes on the following value within an $\vare$-expansion
\begin{equation}
	f'=\dfrac{3}{2}+\dfrac{3d_c\vare}{96+4d_c}+\mathcal{O}(\vare^2)\;.
\end{equation}
If we use an estimate for $\eta$ for $D=2,d_c=1$ from self-consistent calculations \cite{le1992self}, we obtain $f'\approx2.154$.

%\bibliography{refs}
%\bibliographystyle{unsrtnat}

%
\end{document}